\documentclass[pra,nofootinbib,superscriptaddress,floatfix]{revtex4}
\usepackage{graphics}           % for eps figures
\usepackage{bm}                 % for bold math symbols
\usepackage{amsmath}            % for \text and such
\usepackage{amsfonts}
\usepackage{amssymb}
\usepackage{latexsym}           % See if this fixes arXiv problem
\newcommand{\ket}[1]{\mbox{$|#1\rangle$}}
\newcommand{\bra}[1]{\mbox{$\langle#1|$}}
\newcommand{\braket}[2]{\mbox{$\langle#1|#2\rangle$}}
\newcommand{\identity}{\leavevmode\hbox{\small1\kern-3.2pt\normalsize1}}

\begin{document}
%----------------------------%

%%%%%%%%%%%%%%%%%%%%%%%%% Set up the title %%%%%%%%%%%%%%%%%%%%%%%%%%%%%%%%%%
\title{Entanglement in coined quantum walks on regular graphs}

\author{Ivens Carneiro}
\author{Meng Loo}
\author{Xibai Xu}
\affiliation{QOLS, Blackett Laboratory, Imperial College London,
Prince Consort Road, London, SW7 2BW, United Kingdom.}
\author{Mathieu Girerd}
\affiliation{Magist{\`e}re de Physique Fondamentale d' Orsay, Universit\'{e} Paris-Sud, Orsay,   France}
\affiliation{QOLS, Blackett Laboratory, Imperial College London,
Prince Consort Road, London, SW7 2BW, United Kingdom.}
\author{Viv Kendon}
\email{V.Kendon@leeds.ac.uk}
\altaffiliation{Current address: School of Physics and Astronomy, University of Leeds, LS2 9JT, United Kingdom.}
\author{Peter L.~Knight}
%\author{Peter ``Sir Pete'' Knight}
\affiliation{QOLS, Blackett Laboratory, Imperial College London,
Prince Consort Road, London, SW7 2BW, United Kingdom.}

%%% replace with submission date before uploading %%%
%%\date{\today}
\date{Received April 22, 2005, published July 12, 2005\\
xx  New J. Phys. 7 (2005) 156\\
xx  Online at http://www.iop.org/EJ/abstract/1367-2630/7/1/156/\\
xx  doi:10.1088/1367-2630/7/1/000 PII: S1367-2630(05)98776-4\\
xx  Corrections added to figure 2 April 21, 2008}

%%% abstract before title in revtex4 %%%
\begin{abstract}
Quantum walks, both discrete (coined) and continuous time,
form the basis of several recent quantum algorithms.
Here we use numerical simulations to study the properties of discrete,
coined quantum walks.
We investigate the variation in the entanglement between the coin and
the position of the particle by calculating the entropy of
the reduced density matrix of the coin.
We consider both dynamical evolution and asymptotic limits for coins of
dimensions from two to eight on regular graphs.
For low coin dimensions, quantum walks which spread faster (as measured by
the mean square deviation of their distribution from uniform)
also exhibit faster
convergence towards the asymptotic value of the entanglement between
the coin and particle's position.  For high
dimensional coins, the DFT coin operator is more efficient at spreading
than the Grover coin.
We study the entanglement of the coin on regular finite graphs such as
cycles, and also show that on complete bipartite graphs, a quantum walk with a
Grover coin is always periodic with period four.
We generalise the ``glued trees'' graph used by Childs et al.~[STOC, 59, (2003)]
to higher branching rate (fan out) and verify that the scaling with 
branching rate and with tree depth is polynomial.
\end{abstract}

%----------------------- end of title stuff ----------------------------------%

%%%%%%%%%%%%%%%%%%%%%%%%%%%%%%%%%%%%%%%%
%%% Actually produce the title first %%%
\maketitle
%-------------------------------------%

\vspace{-1em} %% Try to keep TOC on first page!

%%%%%% NJP does contents %%%%%%
\tableofcontents
%-----------------------------%

%%%%%%%%%%%%%%%%%%%%%%%% Now begin the text %%%%%%%%%%%%%%%%%%%%%%%%%%%%%%%%%%%

%%%%%%%%%%%%%%%%%%%%%%%%%%%%%%%%%%%%%%%%%%%%%
\section{Introduction}
\label{sec:intro}
%-------------------------------------------%

One of the most important tasks on the theoretical side of quantum
computing is the creation and understanding of quantum algorithms.
The recent presentation of several quantum algorithms based on quantum
versions of random walks is particularly important in this respect,
since they provide a new type of algorithm which can show an
exponential speed up over classical algorithms, to add to those
based on the quantum Fourier transform.
Childs \textit{et al.}~\cite{childs02a} have produced a scheme for a
continuous time quantum walk
that can find its way across a particular ``glued trees'' graph exponentially faster than
any classical algorithm, while Shenvi \textit{et al.}~\cite{shenvi02a} proved
that a discrete quantum walk can reproduce the quadratically faster search times
found with Grover's algorithm for finding a marked item in an unsorted
database.  Generalizations to finding subsets of items have also been developed
\cite{childs03b,magniez05a,ambainis03a}, providing polynomial speed up over
classical algorithms.
For an overview of the development of quantum walks for quantum computing,
see the recent reviews by Kempe \cite{kempe03a} and Ambainis \cite{ambainis04a}.
These results are extremely promising, but still a long way from
the diversity of problems for which classical random walks provide
the best known solutions, such as approximating the permanent of
a matrix \cite{jerrum01a}, finding satisfying assignments to Boolean
expressions ($k$SAT with $k>2$) \cite{schoning99a},
estimating the volume of a convex body \cite{dyer91a},
and graph connectivity \cite{motwani95}.
Classical random walks underpin many standard methods in computational
physics, such as Monte Carlo simulations, 
further motivating the study of quantum walk algorithms.

Like classical random walks, quantum walks come in both discrete time
\cite{aharonov92a,watrous98a,aharonov00a,ambainis01a},
and continuous time \cite{farhi98a} versions.
The discrete and continuous time versions of classical random walks can be
related in a straightforward manner by taking the limit of the
discrete walk as the size of the time step goes to zero.
In the quantum case, the discrete and continuous time walks have
different sized Hilbert spaces so there is no simple limit that relates
the two basic formulations.  There is also an example
of a problem where the algorithmic powers of discrete
and continuous time walks differ.
Spatial search, where there is a cost associated with moving from
one data element to another, can be accomplished faster with a 
discrete time quantum walk, but a continuous time quantum walk
only performs as well for spatial dimensions greater than four \cite{childs03a}.
A continuous time walk with extra degrees of freedom
has also be formulated by Childs and Goldstone \cite{childs04a}
that does correspond to the limit of the discrete time walk and can
perform equally well on spatial search.

Our work in this paper investigates the properties of coins in 
discrete quantum walks.  We follow on from prior work on quantum coins
by \citeauthor{mackay01a} \cite{mackay01a}
and \citeauthor{tregenna03a} \cite{tregenna03a},
broadening the types of graphs studied.
The question of what is particularly quantum in a quantum walk is an
interesting one which has attracted much attention
\cite{knight03a,knight03b,kendon04a}.  In this paper we
address this issue by investigating the evolution of the quantum mechanical
entanglement as the quantum walk progresses.  We quantify the entanglement
between the coin and position for example in a coined walk by using the von
Neumann entropy and show how the entanglement oscillates and approaches
asymptotic values depending on the choice of initial state and coin bias.

The paper is organised as follows:
Walks on infinite lattices are discussed first, starting with
the simple walk on a line in Sec.~\ref{sec:line},
and progressing to walks on lattices in two spatial dimensions
in Sec.~\ref{sec:lattices2D}.
Section \ref{sec:finite} considers walks on finite graphs,
including the $N$-cycle, cycles with diagonals and
complete bipartite graphs.
We then consider quantum walks on the ``glued trees'' graph
of \cite{childs02a}, and generalise it to higher branching rates
in Sec.~\ref{sec:trees}.
Finally, we summarise and conclude in Sec.~\ref{sec:summary}.

%%%%%%%%%%%%%%%%%%%%%%%%%%%%%%%%%%%%%%%%%%%%%%%%
\section{Walk on an infinite line}
\label{sec:line}
%----------------------------------------------%

In a classical random walk on a line, a particle moves either
left or right according to the state of a classical coin where
heads means right and tails means left (or vice versa).
For a quantum version of a random walk, 
the coin is a qubit that can be in a superposition of heads and tails, so
the particle moves left \emph{and} right into a superposition of positions.
This evolution of the walk is governed by a coin
operator that acts on the quantum coin at each step of the walk,
\begin{equation}
\mathbf{C}_2^{(\text{gen})}=\left( \begin{array}{cc}
        \sqrt{\rho}& \sqrt{1-\rho}e^{i\theta}\\
        \sqrt{1-\rho}e^{i\phi}& -\sqrt{\rho}e^{i(\theta+\phi)}
        \end{array} \right),
\label{eq:genH}
\end{equation}
where $0\le\theta,\phi\le\pi$ are arbitrary angles, $0\le\rho\le 1$,
and we have removed an irrelevant global phase so as to leave the
leading diagonal element real.
Equation (\ref{eq:genH}) represents the most general expression
\cite{bach02a}
for a unitary coin operator with two degrees of freedom.
In this expression, the factors $\rho$, and $\theta$, $\phi$ determine the bias
and the phase angles of the coin respectively.
If we set $\rho=0.5$  and $\theta=\phi=0$, the following expression,
called the Hadamard coin operator is obtained:
\begin{equation}
\mathbf{C}_2^{(\text{Had})}=\frac{1}{\sqrt{2}}\left( \begin{array}{cc}
        1 & 1\\
        1 & -1
        \end{array} \right),
\label{eq:had}
\end{equation}
This is an unbiased coin operator, as it chooses the directions left
and right on a line with the same probability. We label as
$\ket{L}$ and $\ket{R}$ the basis states of the coin, 
which can correspond to spin-up and spin-down states respectively. 
We denote the position on the line by $\ket{x}$, so the joint state
of a particle at position $x$ with a coin in state $\ket{L}$ can be written
$\ket{L,x}$.
For the quantum walk on a line,
the phases in the general coin operator ($\theta$ and $\phi$) appear
in the evolution of the walk only in the combination $(\theta+\phi)$,
so as shown by Bach et al~\cite{bach02a}, their effect
is equivalent to varying the phase $\beta$ in the initial coin
state $\ket{\psi_0}$:
\begin{equation}
\ket{\psi_0} = \sqrt{\eta}\ket{L} + \sqrt{1-\eta} e^{i\beta}\ket{R},
\end{equation}
where $\eta$ is the bias in the initial state, and $\beta$ is the
relative phase between the two components.
This leaves only the bias $\rho$ in the coin operator
affecting the outcome of the quantum walk on a line, and, without loss of
generality, we can consider coin operators of the form
\begin{equation}
\mathbf{C}_2^{(\text{bias})}=\left( \begin{array}{cc}
        \sqrt{\rho} & \sqrt{1-\rho}\\
        \sqrt{1-\rho} & -\sqrt{\rho}
        \end{array} \right).
\label{eq:bias}
\end{equation}
After ``flipping'' the coin with the coin operator, the particle moves to 
adjacent positions according the the coin state; this is expressed 
mathematically as a conditional shift operator
\begin{eqnarray}
&&S\ket{L,x} = \ket{L,x-1}\nonumber\\
&&S\ket{R,x} = \ket{R,x+1} .
\label{eq:cshift}
\end{eqnarray}
One complete step of the quantum walk is thus given by the unitary operator
$U = S(C\otimes\identity)$.

%%%%%%%% FIGURE %%%%%%%%
\begin{figure}
    \begin{center}
        %\resizebox{0.65\columnwidth}{!}{\includegraphics{eps/line5000.eps}}
	\resizebox{0.65\columnwidth}{!}{\rotatebox{-90}{\includegraphics{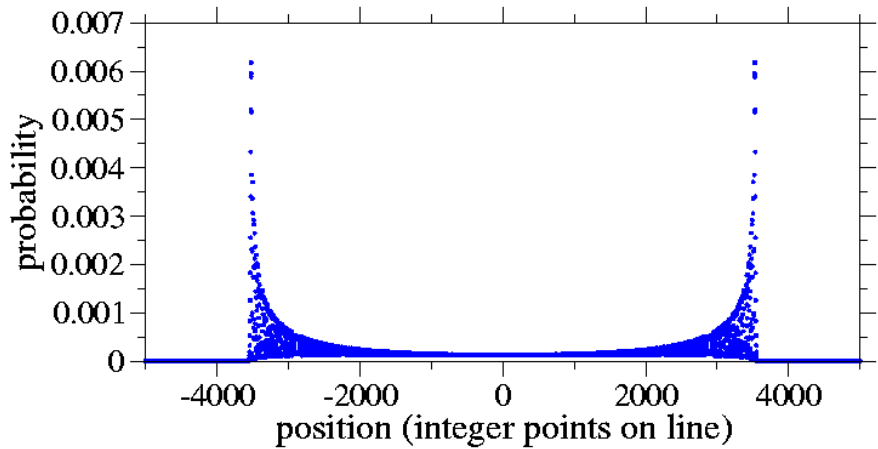}}}
    \end{center}
\caption{Probability distribution for a quantum walk on the line
	after 5000 steps, using a Hadamard coin, equation (\ref{eq:had}),
	and a symmetric initial state $(\ket{R,0}+i\ket{L,0})/\sqrt{2}$.
	Only even positions shown since odd positions are unoccupied.}
\label{fig:line5000}
\end{figure}
The position probability distribution of a quantum walk on a line
is by now well-known, an example
with a Hadamard coin operator, and initial state of 
$\frac{1}{\sqrt{2}}(\ket{R}+i\ket{L})\otimes\ket{0}_x$ after 5000 time steps
is shown in figure \ref{fig:line5000}.

%-------------------------------------------------%
\subsection{Entanglement between coin and position}
\label{ssec:entangle}

Since the quantum walk dynamics are unitary, the system remains in a pure state
and we can use the entropy of the reduced density matrix of the coin
to quantify the entanglement between the coin and the particle's position,
\begin{equation}
E_c(t) = -\sum_j \lambda_j\log_2(\lambda_j)
\label{eq:entdef}
\end{equation}
where $\{\lambda_j\}$ are the eigenvalues of the reduced density matrix
of the coin at time $t$
(in the case of the walk on a line there are just two eigenvalues).
%%%%%%%% FIGURE %%%%%%%%
\begin{figure}
  \begin{minipage}{0.45\columnwidth}
    \begin{center}
%% replace incorrect figure here!  21.4.2008
	%\resizebox{1.0\columnwidth}{!}{\includegraphics{eps/entr_varying_initial-state2.eps}}
	\resizebox{1.0\columnwidth}{!}{\includegraphics{eps/ent200p0.2.eps}}
    \end{center}
  \end{minipage}
\hfill
  \begin{minipage}{0.45\columnwidth}
    \begin{center}
        \resizebox{1.0\columnwidth}{!}{\includegraphics{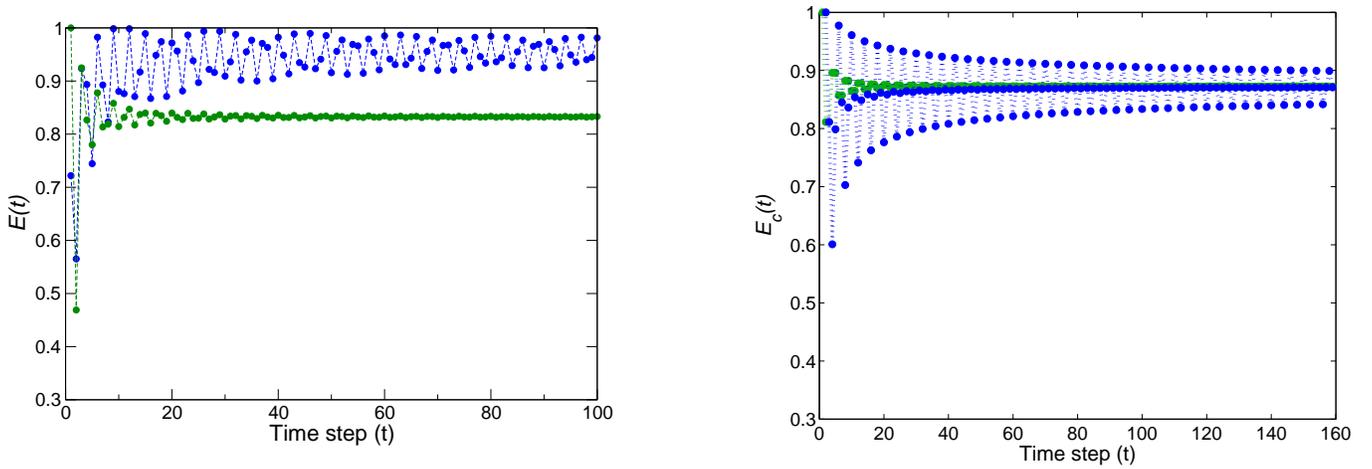}}
    \end{center}
  \end{minipage}
\caption{(Left) Entanglement $E_c(t)$ for a walk on a line with a biased coin,
	equation (\ref{eq:bias}) with $\rho = 0.2$,
	and two different initial states, asymmetric $\ket{R,0}$ (blue) and
	symmetric $(\ket{R,0}+i\ket{L,0})/\sqrt{2}$ (green) \cite{annabestani08a}.
	(Right) Entanglement $E_c(t)$ for a walk on a line with coin operator
	bias $\rho=0.5$ (a Hadamard coin), and the same two initial states.}
\label{fig:entline}
\label{fig:entlinesymasym}
\end{figure}
Figure \ref{fig:entline} shows how $E_c(t)$ varies for different initial
coin states using a coin operator equation (\ref{eq:bias}), with bias $\rho=0.2$.
This shows that the entanglement approaches a limiting value that varies
between zero and one depending on the initial state of the coin.
The rate of convergence to the
limiting value also depends on the initial state of the coin, with symmetric
initial states ($\eta=0.5$) converging fastest (i.e.~oscillations about the
asymptotic value die away fastest).

%----------------------------------------%
\subsection{Limiting value of the entanglement}
\label{ssec:entropylimit}

For the unbiased (Hadamard) coin operator\footnote{
Rold\'{a}n, Knight and Sipe have also studied this case analytically
(unpublished).
}
($\rho=0.5$), whatever initial coin state $\ket{\psi_0}$
is chosen, the asymptotic value of the entanglement
$E_c(t\rightarrow\infty) \longrightarrow E_{\text{Had}}\simeq 0.872$.
Note added: this statement is not correct, the asymptotic value varies with the initial coin state \cite{gattner06a,abal05a}.
However, the rate of convergence is very different for different
initial coin states, with more symmetric initial states converging faster,
see figure \ref{fig:entlinesymasym}.

\begin{figure}
  \begin{minipage}{0.45\columnwidth}
    \begin{center}
	%\resizebox{\columnwidth}{!}{\includegraphics{eps/entropy_rho_t_asym.eps}}
	\resizebox{\columnwidth}{!}{\rotatebox{-90}{\includegraphics{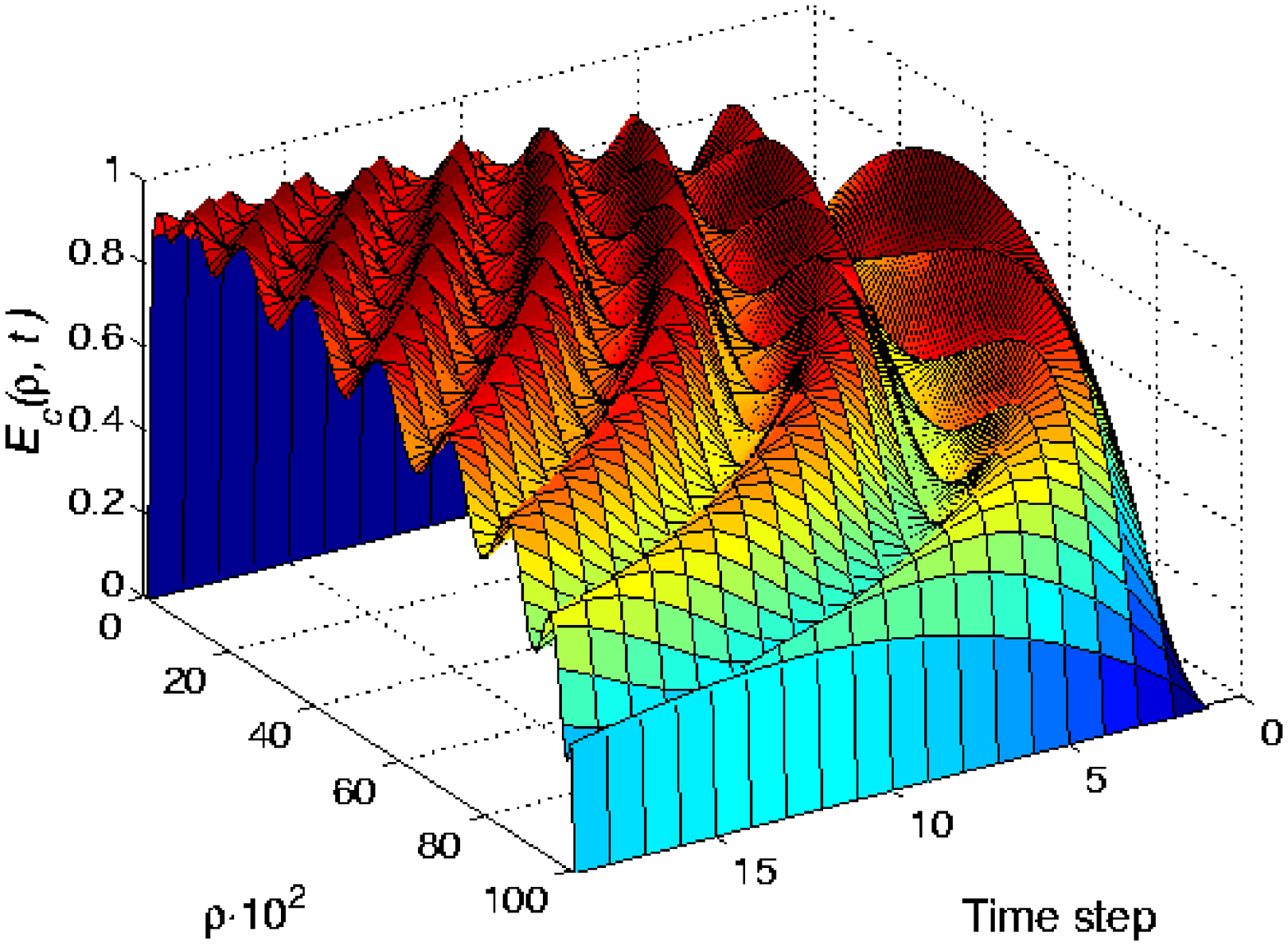}}}
    \end{center}
  \end{minipage}
\hfill
  \begin{minipage}{0.45\columnwidth}
    \begin{center}
	%\resizebox{\columnwidth}{!}{\includegraphics{eps/entropy_rho_t_sym.eps}}
	\resizebox{\columnwidth}{!}{\rotatebox{-90}{\includegraphics{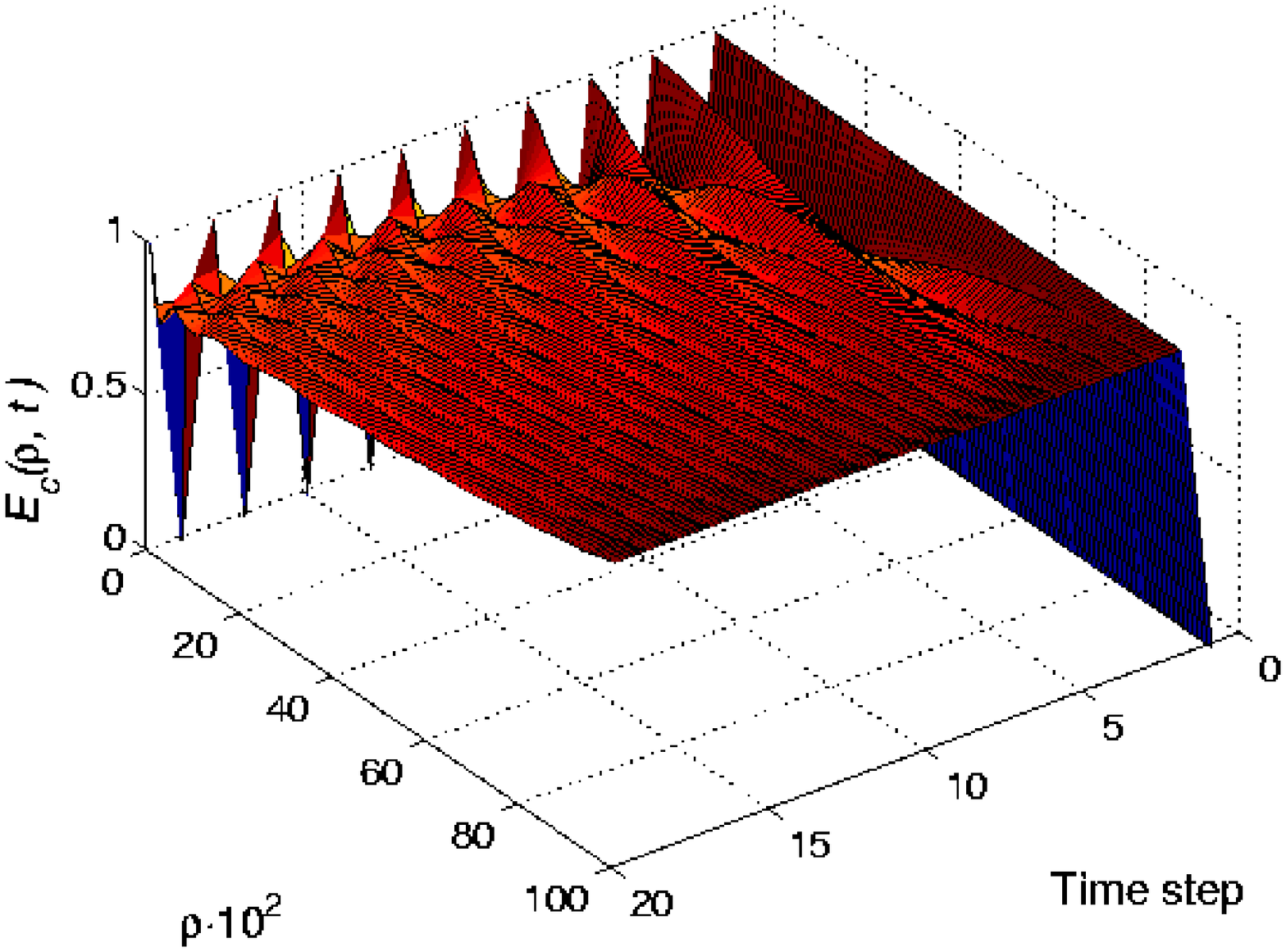}}}
    \end{center}
  \end{minipage}
    \caption{Entanglement $E_c(t)$ for a walk on a line:
	asymmetric initial coin state $\ket{L}$ shown left, and 
	symmetric initial coin state,
	$(\ket{L}+i\ket{R})/\sqrt{2}$ shown right, for
	varying coin bias $\rho$, over 20 time steps.}
\label{fig:rhoEline}
\end{figure}

For biased coin operators, the picture is more complicated. We have 
studied the limiting value of the entanglement for two different initial
state, $\ket{L,0}$ (asymmetric) and $(\ket{L,0}+i\ket{R,0})/\sqrt{2}$
(symmetric), see figure \ref{fig:rhoEline}.
%%%%%%%% FIGURE %%%%%%%%
In the asymmetric case the entanglement converges to a limiting value for
all $\rho$. The limiting value of the entanglement 
$E_{\infty}(\rho)$ increases monotonically from $0$
($\rho=1$) to $1^-$ ($\rho=0^-$) and is
discontinuous at $\rho=0$.  For $\rho=0$ the coin operator becomes the
Pauli spin operator $\sigma_x$ and the entanglement is zero for all time steps,
showing the coin and particle remain disentangled.
In the symmetric case the entanglement
converges to a limiting value for all except the extreme case of
$\rho=0$ where the entanglement oscillates between the minimum and
maximum values $0$ and $1$, as can easily be verified analytically.
The asymptotic value of the entanglement $E_{\infty}(\rho)$
increases to $1$ as $\rho$ increases.
The other variable factor is the period of the oscillations
about the convergent value.
In both cases, except for $\rho=0$ this period increases to
infinity as $\rho\rightarrow1$.
As noted for the unbiased (Hadamard) coin, the rate of convergence to
the asymptotic entanglement is faster for the symmetric initial coin state.

%----------------------------------------%
\subsection{Rate of convergence}
\label{ssec:converge}

\begin{figure}
    \begin{center}
	\resizebox{0.6\columnwidth}{!}{\includegraphics{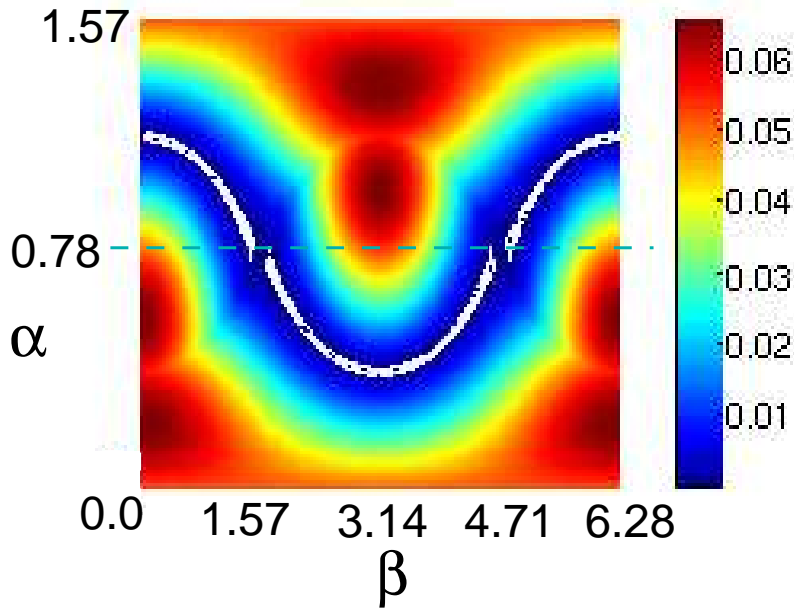}}
    \end{center}
    \caption{Amplitude of oscillations of entanglement
	versus initial coin state parameters $\alpha$ and $\beta$,
	for a Hadamard coin operator applied for 200 time steps.}
    \label{fig:alphabeta}
\end{figure}

In order to quantify the rate of convergence of the entanglement to its
limiting value, we considered the magnitude of the entanglement at
a fixed time while varying the initial state.
It is convenient to write the initial coin state as
\begin{equation}
\ket{\psi_0}=\cos(\alpha)\ket{R} + \sin(\alpha)e^{i\beta}\ket{L},
\end{equation}
where $\eta=\cos^2(\alpha)$.
Figure \ref{fig:alphabeta} shows how the amplitude varies with both
$\alpha$ and $\beta$.
First consider the case where $\alpha=\frac{\pi}{4}$ for $\eta=0.5$ i.~e.~a
symmetric initial state.
We find that the entanglement is proportional to $\cos^2(\beta)$ along 
the dotted line in figure \ref{fig:alphabeta}.
We can also fit a formula for the minimum, 
\begin{equation}
\alpha=\frac{\pi}{8}\cos(\beta) + \frac{\pi}{4}.
\end{equation} 
%%%%%%%% FIGURE %%%%%%%%
This is the white line in the blue region in figure \ref{fig:alphabeta} where 
the entanglement oscillations are smallest, i.e., fastest convergence.

To summarise the results for the quantum walk on a line,
we find the various behaviours of the entanglement are governed as follows.
The asymptotic value $E_{\infty}$ reached by the entanglement 
is a function of both the coin bias $\rho$ and the initial
state $\ket{\psi_0}$.
For a fixed number of time steps $t$, the period of oscillation
of the entanglement $E_c(t)$ around $E_{\infty}$ is a
function of $\rho$ only. 
For the special case of $\rho=0.5$,
an unbiased coin, $E_{\infty}=E_{Had}$
has the same value of $0.872\dots$ for all choices of initial coin
state $\ket{\psi_0}$.

%----------------------------------------%
\section{Lattices in two spatial dimensions}
\label{sec:lattices2D}

%----------------------------------------%
\subsection{Higher dimensional coins}
\label{ssec:coins}

For lattices with more than two edges meeting at each vertex,
there is a far wider range of unitary coin operators, since
the coin must now have as many degrees of freedom as there are choices of path.
Since the range of higher dimensional coin operators is too
large for systematic numerical study, for the remainder of this
paper we concentrate on two natural choices.
The Grover operator was first introduced by
\citeauthor{moore01a}~\cite{moore01a}
in their study of quantum walks on the hypercube.  
Based on Grover's diffusion operator, it has elements
$[C^{(\text{G})}_d]_{i,j} = \frac{2}{d} - \delta_{ij}$, i.e.,
\begin{equation}
  \mathbf{C}_d^{(G)}=
  \left(
    \begin{array}{ccccc}
    2/d-1&2/d&\ldots&2/d\\
    2/d&2/d-1&\ldots&2/d\\
    \vdots&\vdots& &\vdots\\
    2/d&2/d&\ldots&2/d-1\\
    \end{array}
  \right).
\label{eq:grovd}
\end{equation}
For example, the $d=3$ case is 
\begin{equation}
\mathbf{C}_3^{(G)}=\frac{1}{3}\left( \begin{array}{rrr}
        -1 & 2 & 2\\
         2 &-1 & 2\\
         2 & 2 &-1\\
        \end{array} \right).
\label{eq:grov3}
\end{equation}
Except in the $d=4$ case, the Grover coin is biased,
since the incoming direction (corresponding to the diagonal entry)
is treated differently from the outgoing directions.
However, it is symmetric under interchange of any outgoing coin directions,
and is in fact the symmetric unitary operator furthest from the identity.
The $d=4$ Grover coin is the only unbiased Grover coin since all the entries
are $\pm\frac{1}{2}$
\begin{equation}
\mathbf{C}_4^{(G)}=\frac{1}{2}\left( \begin{array}{rrrr}
        -1 & 1 & 1 & 1\\
         1 &-1 & 1 & 1\\
         1 & 1 &-1 & 1\\
         1 & 1 & 1 &-1\\
        \end{array} \right).
\label{eq:grov4}
\end{equation}

The DFT (discrete Fourier transform) coin is unbiased for all $d$,
but asymmetric in that you cannot interchange the
labels on the directions without changing the coin operator: each direction
acquires its own phase shift.
For $d=3$, it looks like
\begin{equation}
\mathbf{C}_3^{(D)}=\frac{1}{\sqrt{3}}\left( \begin{array}{ccc}
        1 & 1 & 1\\
        1 & \omega_3 & \omega^2_3\\
        1 & \omega^2_3 & \omega_3\\
        \end{array} \right),
\label{eq:DFT3}
\end{equation}
where $\omega_3=e^{2i\pi/3}$ and $\omega^2_3=e^{-2i\pi/3}$ are the
complex cube roots of unity.
The $d$-dimensional DFT coin can be written
\begin{equation}
  \mathbf{C}_d^{(D)}=\frac{1}{\sqrt{d}}
  \left(
    \begin{array}{ccccc}
    1&1&1&\ldots&1\\
    1&\omega&\omega^2&\ldots&\omega^{d-1}\\
    1&\omega^2&\omega^4&\ldots&\omega^{2(d-1)}\\
    \vdots&\vdots&\vdots& &\vdots\\
    1&\omega^{d-1}&\omega^{2(d-1)}&\ldots&\omega^{(d-1)^2}\\
    \end{array}
  \right),             
\label{eq:DFTd}
\end{equation}
where $\omega$ is the complex $d$'th root of unity.

%----------------------------------------%
\subsection{Cartesian lattice}
\label{ssec:lattices2D}

For a 2-dimensional Cartesian grid, there are four edges meeting
at each lattice site, so a $d=4$ dimensional coin is required.
The quantum walk is a generalisation of the walk on a line.
We tested both Grover and DFT coins ($d=4$ versions) and found a
similar range of behaviours for the entanglement between the coin and
the position as for the walk on a line,
only compounded by having twice as many directions.  So, for
example, the period of the oscillations about the asymptotic value is
now a more complicated pattern of two frequencies.
%%%%%%%% FIGURE %%%%%%%%
\begin{figure}
    \begin{center}
    \begin{minipage}{0.45\columnwidth}
	\resizebox{1.0\columnwidth}{!}{\includegraphics{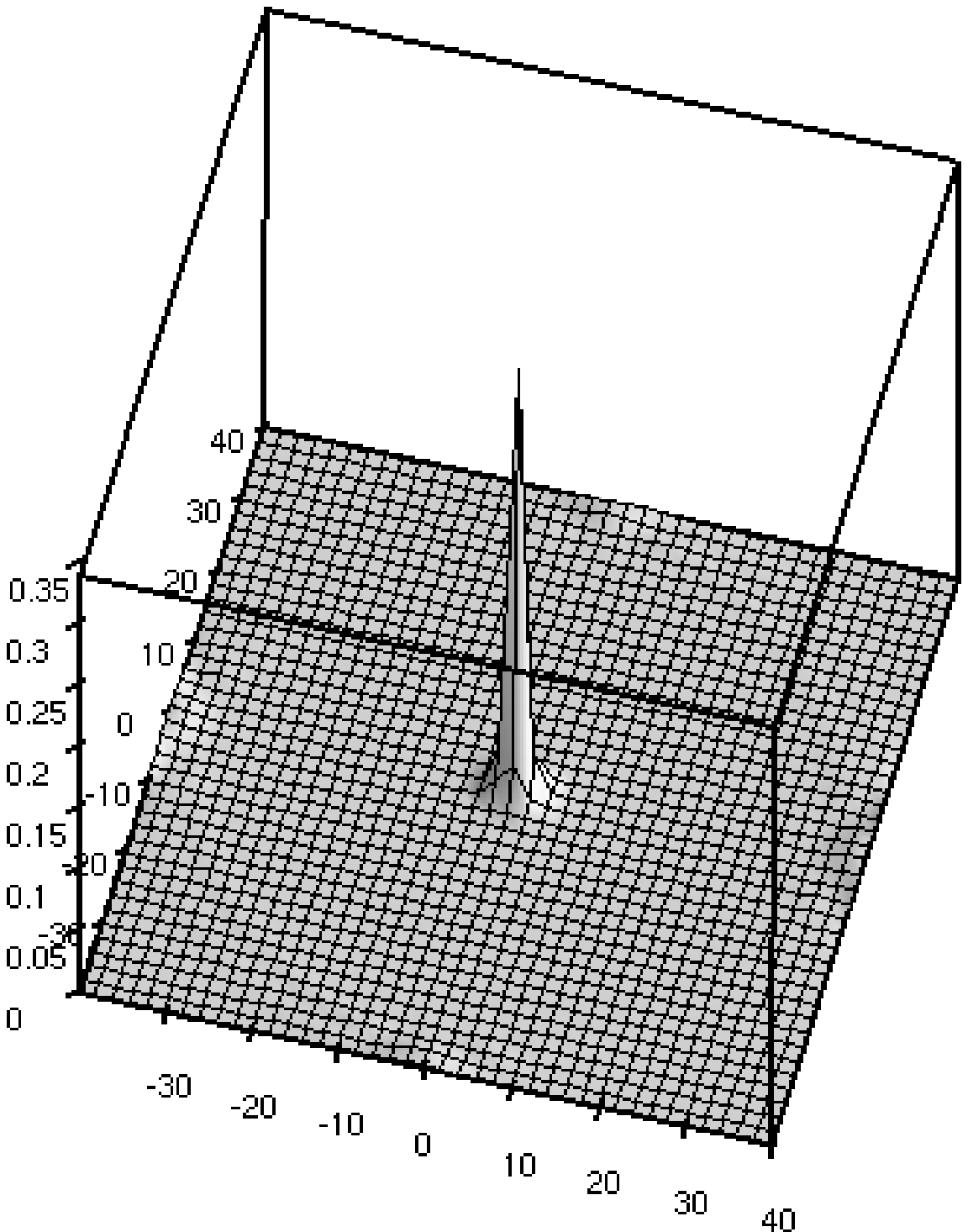}}
    \end{minipage}
    \hfill
    \begin{minipage}{0.45\columnwidth}
	\resizebox{1.0\columnwidth}{!}{\includegraphics{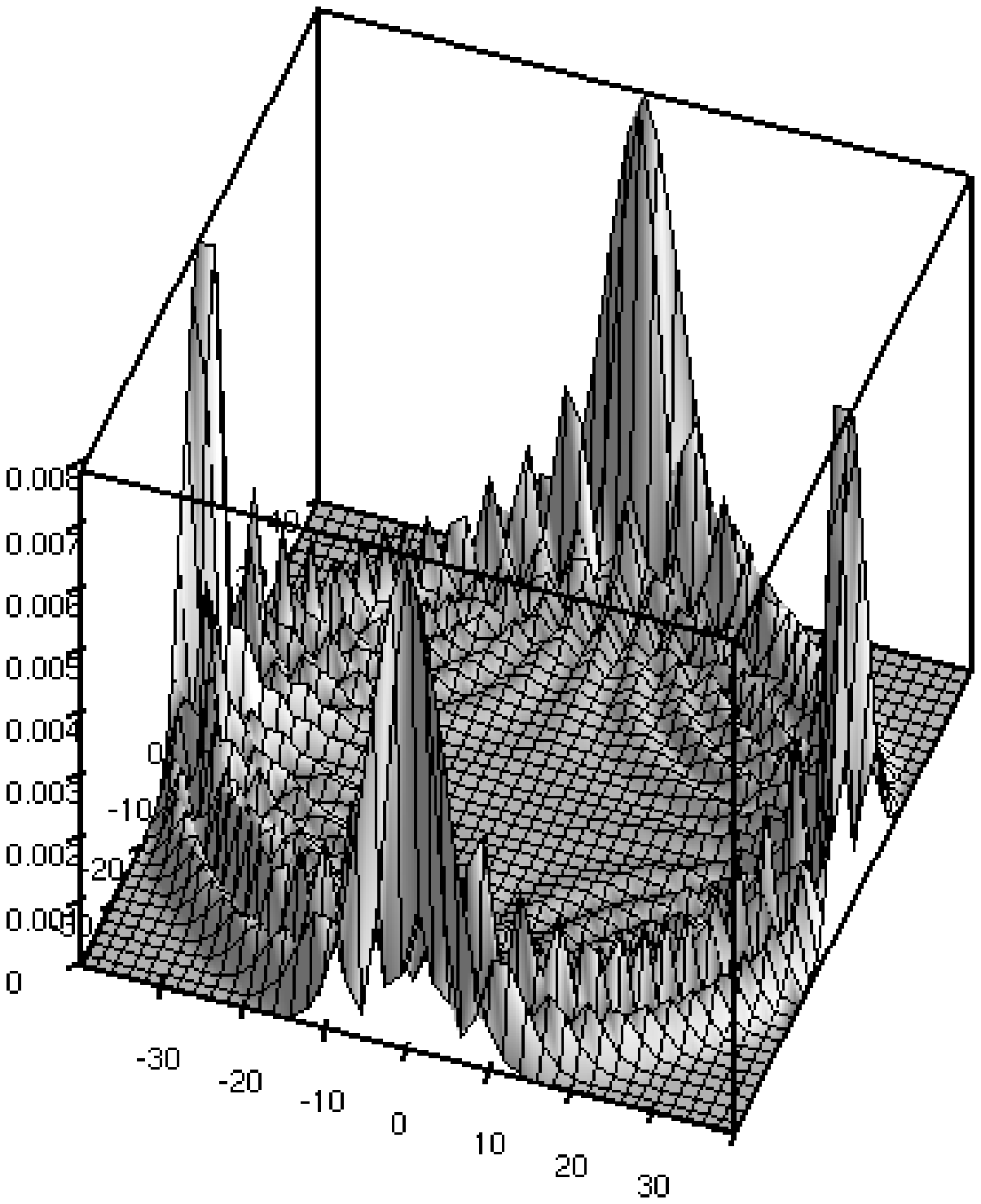}}
    \end{minipage}
    \end{center}
\caption{Typical (left) and ring-like (right) spreading distribution for a Grover coin on a 2-dimensional Cartesian lattice (from \cite{tregenna03a}).
Axes represent position ($x$ and $y$) with the $z$ axes indicating the
probability of finding the particle at that position.
%%% Remove below for ArXiv version!!!
%\textcolor[named]{Red}{(Note to referees and editors:
%these figures are included for your
%convenience while reviewing the paper; since \cite{tregenna03a} is
%published in NJP, we suggest replacing them in the final version
%with a suitable link to the previously published figures.)}
%%% end remove for ArXiv version!!!
}
\label{fig:spikering}
\end{figure}
We looked for a correlation between the rate of convergence of the
entanglement and the degree to which the quantum walk spreads out over the
lattice.  Most choices of initial state for the Grover and DFT
coin operators produce
a high probability of finding the particle on or near the starting point,
with only one special initial state giving a high rate of spreading,
compare the two distributions in figure \ref{fig:spikering}, taken from
\cite{tregenna03a}.
Spreading is a property of random walks that can be useful for efficient,
uniform sampling, compare \cite{kendon02c}.
The entanglement converges much faster for the quantum walk that spreads out
in the ring, see figure \ref{fig:2Dgrovent}.
The entanglement between the coin and the position
thus provides a way to monitor the progress and character of the walk.
%%%%%%%% FIGURE %%%%%%%%
\begin{figure}
    \begin{center}
	\resizebox{0.5\columnwidth}{!}{\includegraphics{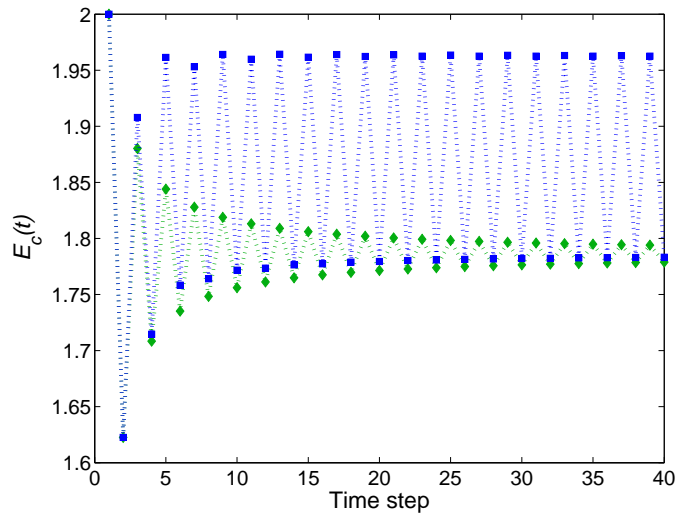}}
    \end{center}
\caption{Entanglement $E_c(t)$ for the spike (blue) and ring (green)
	distributions on a 2D Cartesian lattice, using a Grover coin.
	A DFT coin operator produces a very similar result.}
%% \textcolor[named]{Red}{
%% \textbf{I don't think we have any idea what parameters the two different lines 
%% correspond to here...maybe it doesn't matter?}}
\label{fig:2Dgrovent}
\end{figure}

%\clearpage

%----------------------------------------%
\subsection{Triangular lattices}
\label{ssec:triangular}

\begin{figure}
    \begin{center}
	\resizebox{0.5\columnwidth}{!}{\includegraphics{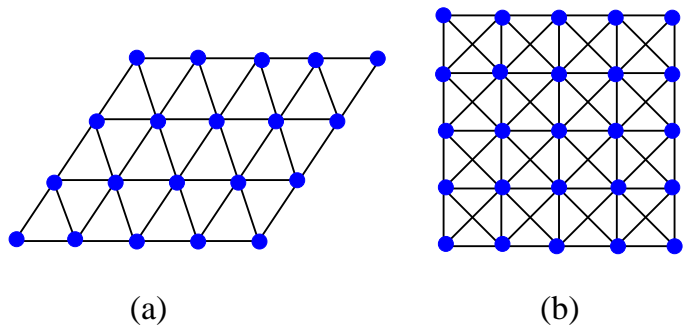}}
    \end{center}
\caption{Triangular lattices with (a) $d=6$ and (b) $d=8$.}
\label{fig:grids}
\end{figure}

Higher dimensional lattices that lie in a plane (two spatial
dimensions) can be constructed in a number of ways.  We studied two
examples: a tessellation of equilateral triangles produces a lattice
with $d=6$;
and adding diagonals to a Cartesian grid, makes
a ``first and second nearest neighbours'' lattice with $d=8$.
These are illustrated in figure \ref{fig:grids}.
%%%%%%%% FIGURE %%%%%%%%

We tested a number of different initial states with both Grover and DFT coins,
again looking at correlations between the amount that a walk starting from
the centre of the grid spreads and the oscillations in the entanglement.
To quantify the spread we studied the mean square deviation of the probability
distribution from the uniform probability distribution
\begin{equation}
\overline{\Delta P}^2(t) = \sum_r \left( |\psi(r,t)|^2  - \frac{1}{N(t)} \right)^2,
\end{equation}
where $r$ is a lattice site in the set it is possible to reach
after $t$ steps, and $N(t)$ is the number of such lattice
sites (so $1/N(t)$ is the average probability per site).
If the walk is spread out evenly over the lattice then $\overline{\Delta P}^2(t)$
will be small, whereas if it is concentrated on parts of the lattice,
$\overline{\Delta P}^2(t)$ will be larger.

\begin{figure}
  \begin{minipage}{0.45\columnwidth}
    \begin{center}
        %\resizebox{\columnwidth}{!}{\includegraphics{eps/prob-grid6-100-symm-i.eps}}
	\resizebox{\columnwidth}{!}{\includegraphics{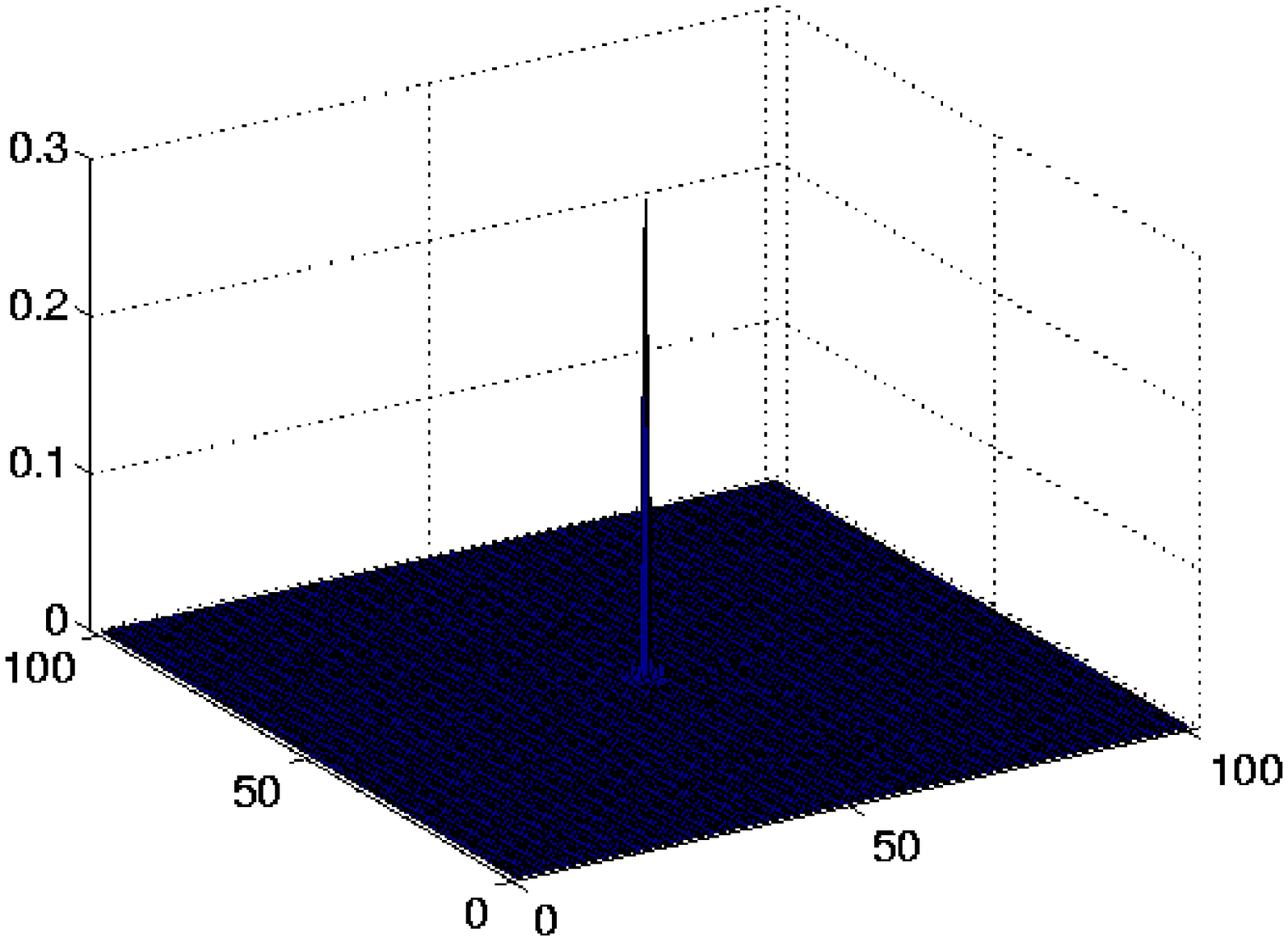}}
    \end{center}
  \end{minipage}
  \begin{minipage}{0.45\columnwidth}
    \begin{center}
        %\resizebox{\columnwidth}{!}{\includegraphics{eps/prob-grid6-100-symm.eps}}
	\resizebox{\columnwidth}{!}{\includegraphics{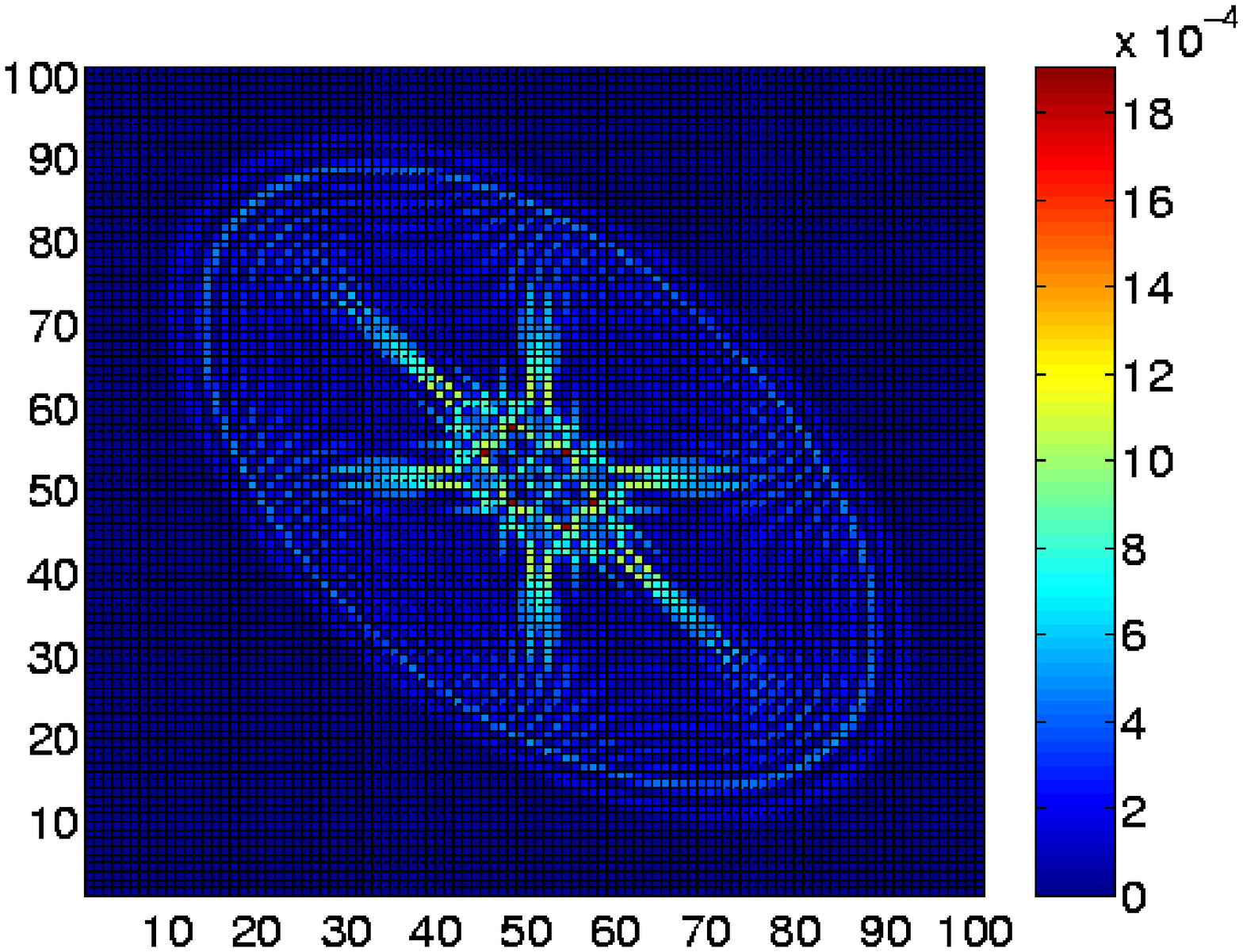}}
    \end{center}
  \end{minipage}
\caption{Probability distribution after 49 steps of a quantum walk on a $d=6$
	grid using a Grover coin operator:
	(Left) starting with coin state
	$\frac{1}{\sqrt{d}}\sum_{i=1}^d|i\rangle$, at the point (50,50), and
	(Right) starting with coin state
	$\frac{1}{\sqrt{d}}\sum_{i=1}^{d/2}|i\rangle -
	\frac{1}{\sqrt{d}}\sum_{i=d/2+1}^d|i\rangle$.
	Axes represent position in the $xy$-plane and probability as
	in figure \ref{fig:spikering}.}
%\textcolor[named]{Red}{\textbf{No source for figure...OK?}}
\label{fig:triangleprob}
\end{figure}

\begin{figure}
  \begin{minipage}{0.45\columnwidth}
    \begin{center}
        \resizebox{\columnwidth}{!}{\includegraphics{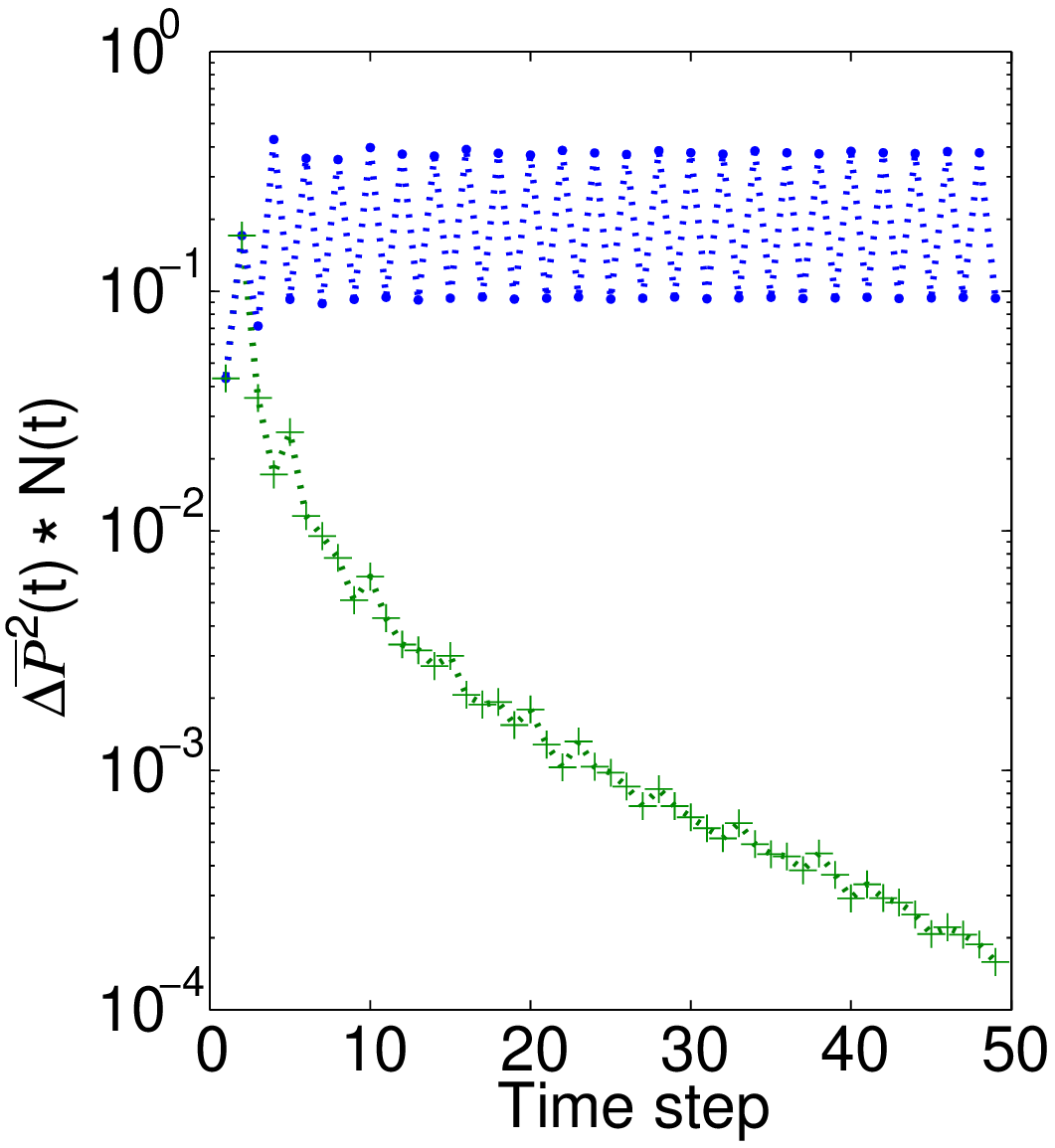}}
    \end{center}
  \end{minipage}
  \begin{minipage}{0.45\columnwidth}
    \begin{center}
        \resizebox{\columnwidth}{!}{\includegraphics{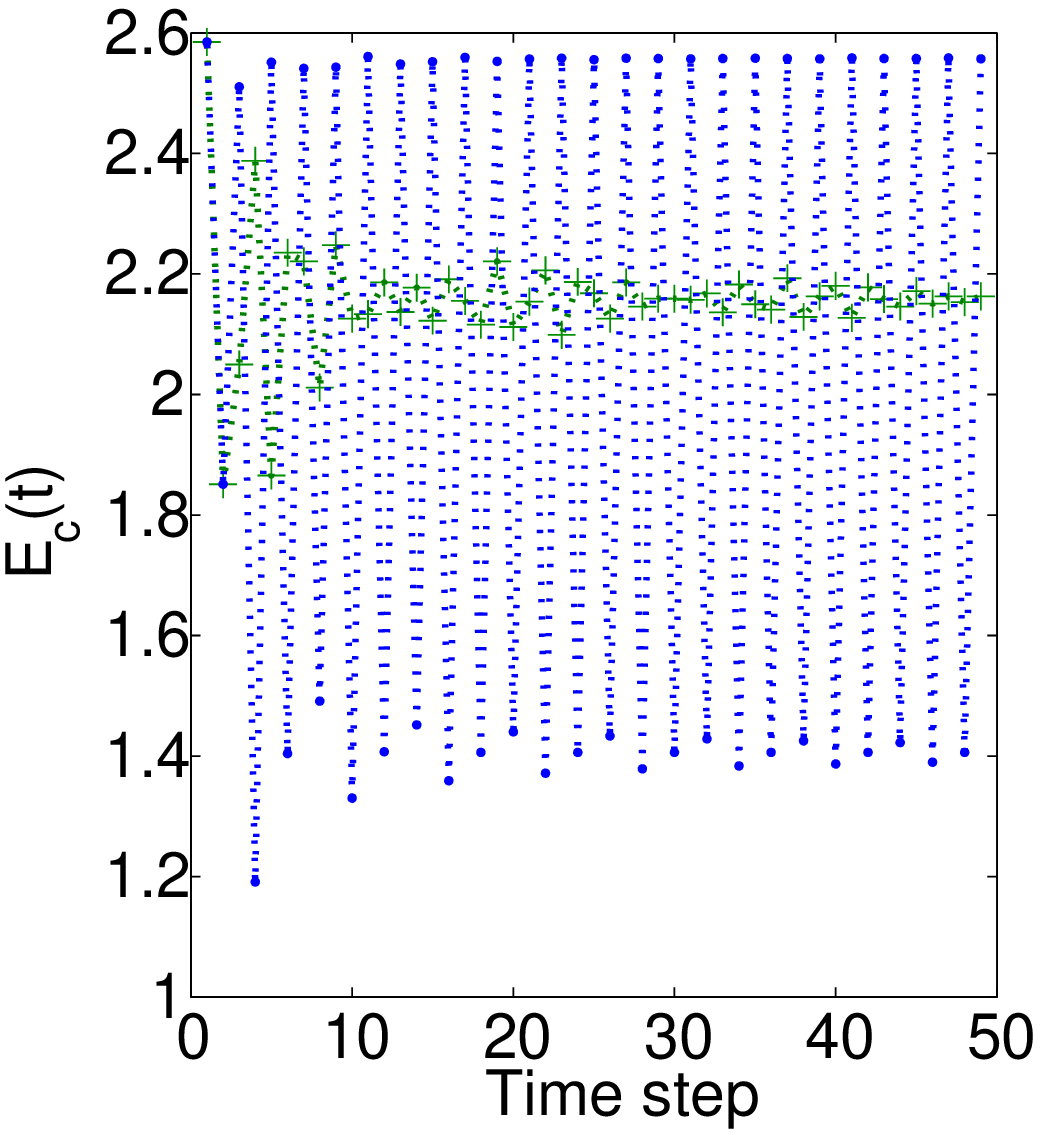}}
    \end{center}
  \end{minipage}
\caption{Spreading as indicated by the mean square deviation from the 
	uniform distribution $\overline{\Delta P}^2(t)$ (left) and
	entanglement $E_c(t)$ (right) for the
	cases shown in figure \ref{fig:triangleprob} spike (blue)
	and ring (green).}
%%\textcolor[named]{Red}{\textbf{No source for figure...OK?}}
\label{fig:trianglenti6}
\end{figure}

For the both the $d=6$ and $d=8$ grids, as on the rectangular grid,
the Grover coin can produce two kinds of behaviour:
fast spreading distributions and distributions 
concentrated nearly all close to the origin, depending on the choice
of initial states.
The amplitudes of the oscillations in the entanglement 
decrease quickly for fast spreading but only slowly for the distributions
stuck near the starting point.
This is illustrates in figures \ref{fig:triangleprob}-\ref{fig:trianglenti6}. 
%%%%%%%% FIGURE %%%%%%%%
%%%%%%%% FIGURE %%%%%%%%

In contrast to the Grover coin operator, the DFT coin produces
good spreading for almost all chosen initial states and the
entanglement converges faster for these cases.
To investigate the correlation between the entanglement and the spreading
of the walk more thoroughly, we wrote the initial coin state as
\begin{equation}
  |\psi_0\rangle=\sum_{j=1}^de^{i\phi_j}|j\rangle.
\end{equation}
For $d=6$,
we fixed one of the phases and varied the other $d-1$ over all
the permutations, allowing repetitions of $\phi_j=\pi j/d$, 
($j=1,\ldots,d$, total of $6^6$ values, thus) 
to see the interference effects on the spread.
For the Grover coin operator, most of these states give a large 
$\overline{\Delta P}^2(t)$ while for the DFT most give a small 
$\overline{\Delta P}^2(t)$: the average (over these initial states) 
after $t=15$ steps is $\langle \overline{\Delta P}^2(t)\rangle=0.0501$ for the Grover and 
$\langle \overline{\Delta P}^2(t)\rangle=0.0015$ for the DFT walk.
For the minimum $\overline{\Delta P}^2(t)$ in this set,
the entanglement oscillations decay fastest,
as seen in figures \ref{fig:trianglenti6} (for $d=6$)
and \ref{fig:trianglentsym} (for $d=8$).
The minimum values of $\overline{\Delta P}^2(t)$  occur for the initial states 
$\{\phi_j\}=(0,0,0,0,0,0)$ for the Grover coin operator and
$\{\phi_j\}=(0,0,4,1,1,5)$ for the DFT coin operator.

For $d=8$, we fixed one of the phases and varied the other $d-1$ over 
only the following subsets of all the
arrays of $\phi_j=\pi j/d$: (i) $j\in{0,d-1}$,
$\{\phi_j\}=(0,0,0,0,0,0,0,0), 
(0,0,0,0,0,0,0,d-1),\ldots,(d-1,d-1,\ldots,d-1,d-1,0)$; 
(ii) permutations of $(0,1,2,\ldots,d-1)$ (without repetitions)
to see the interference effects on the spread.
Although we didn't vary over as large a range as in the $d=6$ case,
we could observe a similar behaviour: better spread for DFT against Grover, 
entropy converging quickly for better spread.

The average (over these initial states) 
after $t=15$ steps is $\langle \overline{\Delta P}^2(t)\rangle=0.0508$
for the Grover and $\langle \overline{\Delta P}^2(t)\rangle=0.0027$ for the DFT walk. 
(For the $d=8$ grid $N(t)=(2t+1)^2$, since there are diagonals 
everywhere).
The minimum values of $N(t)\overline{\Delta P}^2(t)$ occur for the initial states 
$\{\phi_j\}=(0,0,0,0,0,0,0,0)$ for the Grover coin operator
and $\{\phi_j\}=(0,7,7,7,7,0,0,7)$ for the DFT coin operator. 
%%%%%%%% FIGURE %%%%%%%%
\begin{figure}
    \begin{center}
        %\resizebox{0.5\columnwidth}{!}{\includegraphics{eps/prob-grid8-80-symm.eps}}
	\resizebox{0.5\columnwidth}{!}{\includegraphics{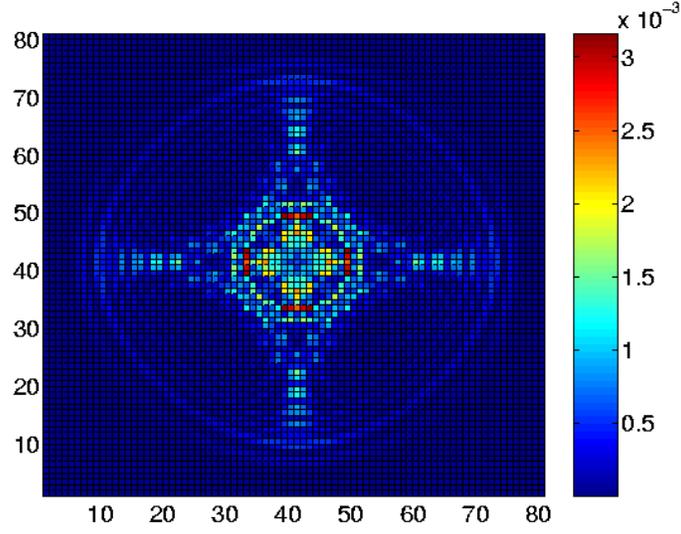}}
    \end{center}
\caption{Grover coin on $d=8$ grid, after 39 time steps, start with coin state
	$\frac{1}{\sqrt{d}}\sum_{i=1}^{d/2}|i\rangle -
  \frac{1}{\sqrt{d}}\sum_{i=d/2+1}^d|i\rangle$.
	Axes represent position in the $xy$-plane and probability as
	in figure \ref{fig:spikering}.}
%%\textcolor[named]{Red}{\textbf{No source for figure...OK?}}
\label{fig:triangleprobsym}
\end{figure}
%%
%%%%%%%% FIGURE %%%%%%%%
\begin{figure}
  \begin{minipage}{0.45\columnwidth}
    \begin{center}
        \resizebox{\columnwidth}{!}{\includegraphics{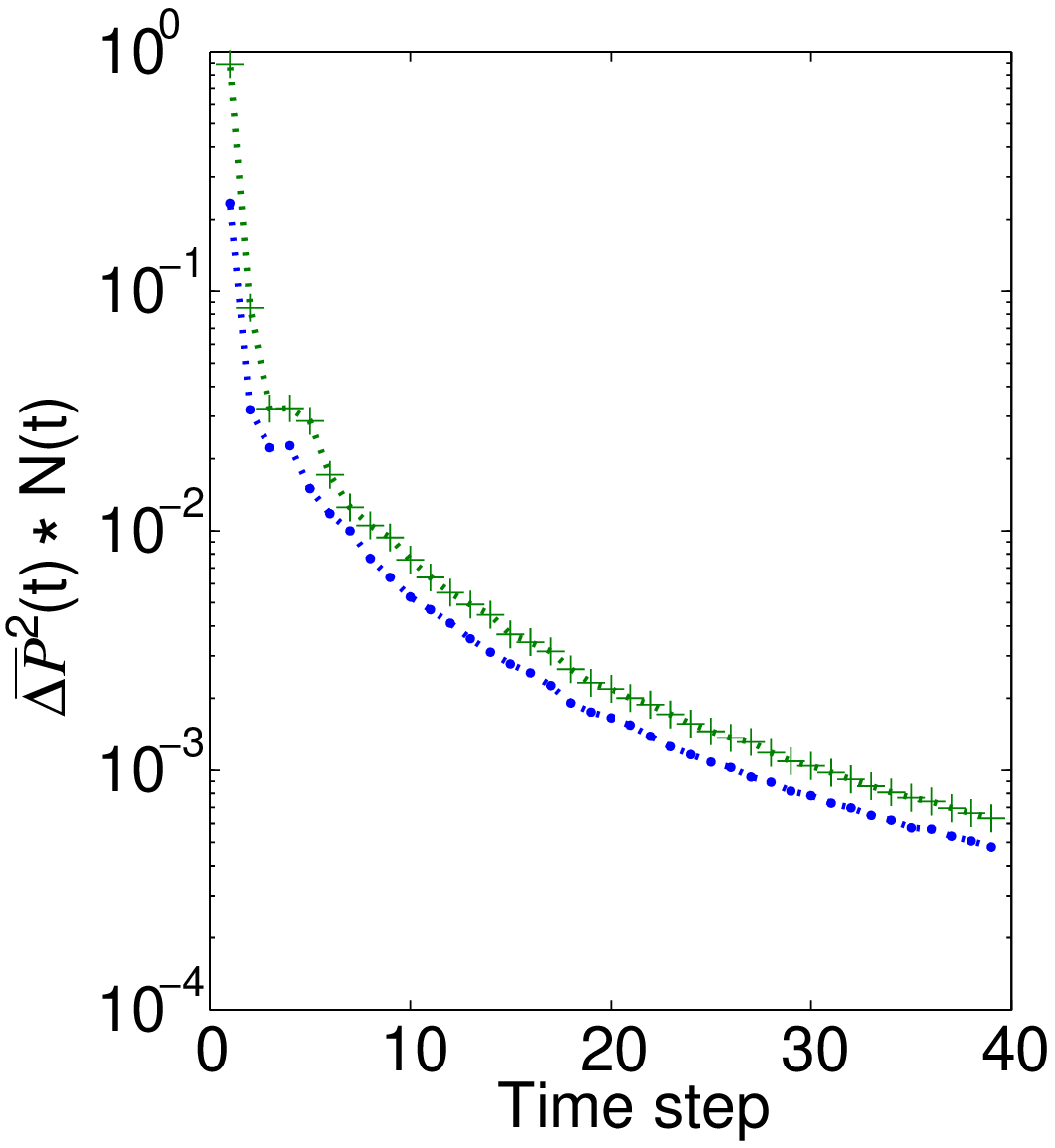}}
    \end{center}
  \end{minipage}
  \begin{minipage}{0.45\columnwidth}
    \begin{center}
        \resizebox{\columnwidth}{!}{\includegraphics{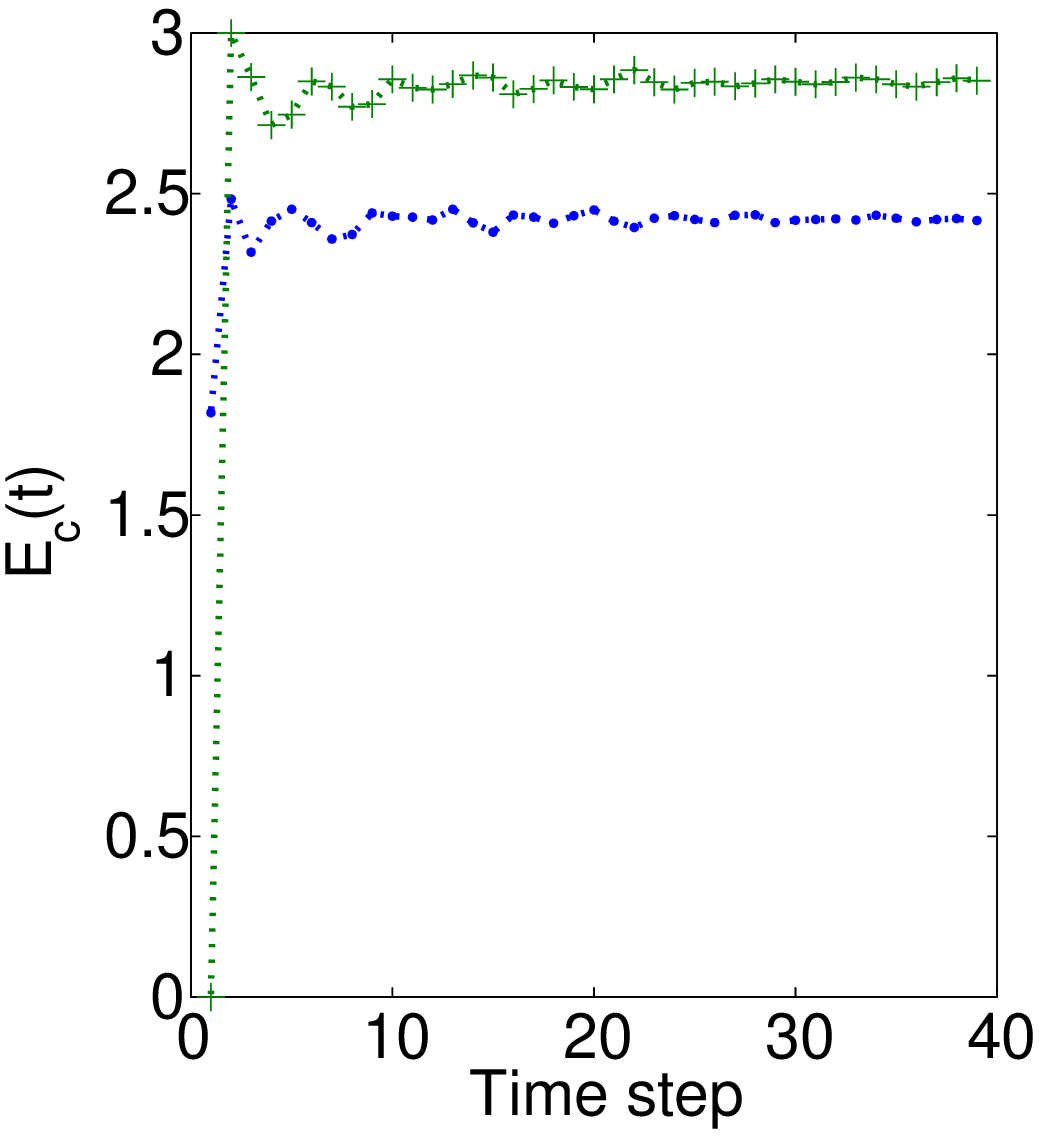}}
    \end{center}
  \end{minipage}
\caption{Spreading as indicating by the mean square deviation from the
	uniform distribution $\overline{\Delta P}^2(t)$ (left) and
	entanglement $E_c(t)$ (right) for the DFT coin operator on $d=8$ grid,
	after 39 time steps, starting with coin states
	$\frac{1}{\sqrt{d}}\sum_{i=1}^d|i\rangle$ 
	(blue) and $\frac{1}{\sqrt{d}}\sum_{i=1}^{d/2}|i\rangle + i
	\frac{1}{\sqrt{d}}\sum_{i=d/2+1}^d|i\rangle$ (green).}
%%\textcolor[named]{Red}{\textbf{No source for figure...OK?}}
\label{fig:trianglentsym}
\end{figure}
For the $d=8$ grid we also tested Hadamard coins
(a simple 2D Hadamard coin for each of the 4 pairs of opposite directions)
and found they displayed similar properties to the DFT coin.

Thus, for all tested coin operators and coin initial states,
we have verified that those walks with good spreading
also show small amplitudes of oscillation in the entanglement of the coin,
pointing to a quick convergence.
In particular, the DFT and Hadamard coin operators produce faster spreading
than the Grover coin for most choices of initial state on these
lattices of higher degree.
This can be explained by noticing that for $d>4$, the Grover
coin is biased so that it favours returning along the edge that it arrived
from.  This will tend to reduce its spreading power.
The DFT coin is unbiased, so its spreading power is affected only by how
the different phases cause interference effects.  For larger $d$, there
are more different phases, so less opportunities for them to all cancel out.
The Hadamard coin is also unbiased, and it does not mix between
the different orientations of the pairs of edges, so on these triangular
lattices it produces a spreading equivalent to the spreading on a line.

%%\clearpage

%%%%%%%%%%%%%%%%%%%%%%%%%%%%%%%%%%%%%%%%%%%%%%%%
\section{Walks on finite regular graphs}
\label{sec:finite}
%----------------------------------------------%

We now turn to quantum walks on graphs with a fixed number of vertices
so the walk is bounded and the notion of spreading is no longer the
relevant property.
Quantum walks on finite graphs were first investigated by
Aharonov et al~\cite{aharonov00a}, who showed that 
while the instantaneous distribution of a quantum walk on these graphs
does not converge (being unitary and reversible), a suitably defined
time-averaged distribution always converges, though
this distribution need not be uniform (in the classical case the limiting
distribution is always uniform).
The interesting questions are thus how fast the quantum walk converges to the
limiting time-averaged distribution, and whether the distribution is
uniform.
We are also interested in whether the walk shows
periodic behaviour in the instantaneous distribution
\cite{tregenna03a}, and if so, under what conditions.

%----------------------------------------------%
\subsection{$N$-cycles}
\label{ssec:ncycles}

\begin{figure}
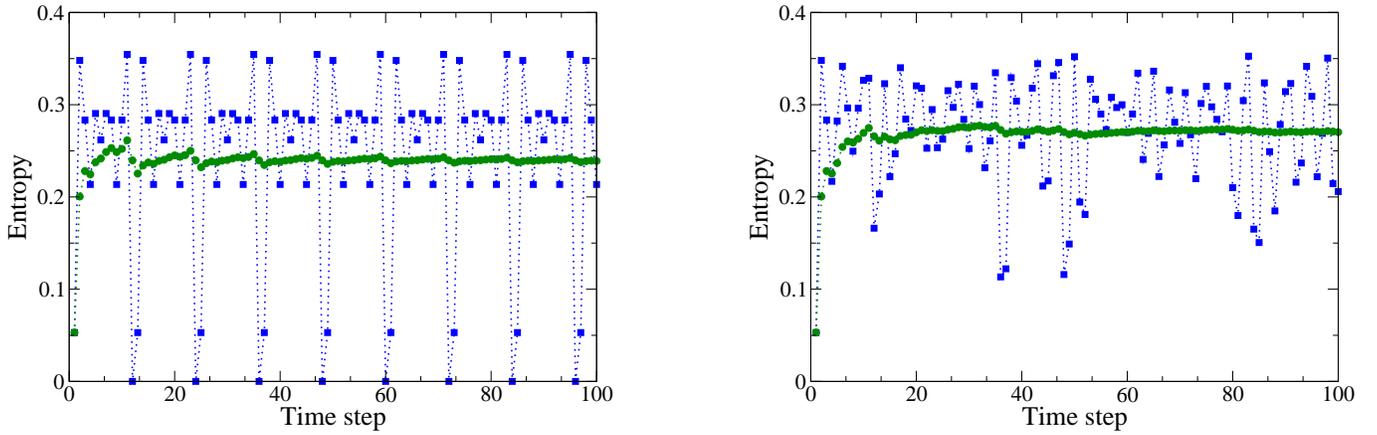

  \begin{minipage}{0.45\columnwidth}
    \begin{center}
        \resizebox{\columnwidth}{!}{\includegraphics{eps/entcycle8.eps}}
    \end{center}
  \end{minipage}
\hfill
  \begin{minipage}{0.45\columnwidth}
    \begin{center}
        \resizebox{\columnwidth}{!}{\includegraphics{eps/entcycle16.eps}}
    \end{center}
  \end{minipage}
\caption{(Left) Entropy versus time step for eight-cycle with 
	symmetric initial coin state.
	(Right) Entropy versus time step for 16-cycle with 
	symmetric initial coin state. Both instantaneous (blue)
	and time averaged (green) values of the entropy are shown.}
\label{fig:8+16cycle}
\end{figure}

\begin{figure}
    \begin{center}
        \resizebox{0.65\columnwidth}{!}{\includegraphics{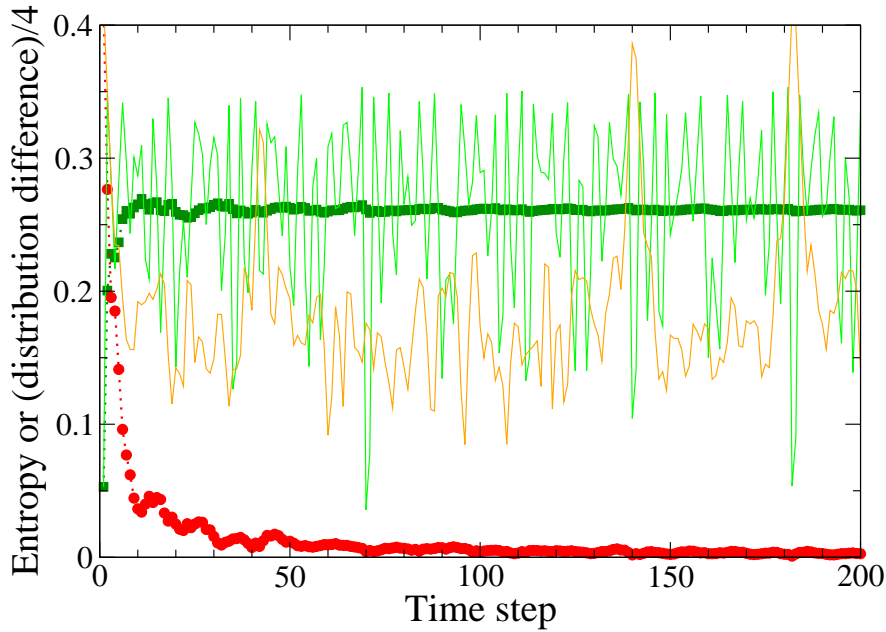}}
    \end{center}
\caption{Comparison of time-averaged entropy (thick green) with convergence to
time-averaged limiting (uniform in this case) distribution (thick red)
for a seven-cycle with symmetric initial coin state and Hadamard coin operator.
The difference between the time-averaged distribution and the limiting
distribution is scaled by a factor of 0.25 to make the range of fluctuations
roughly comparable with the entropy, as shown by the instantaneous entropy
and distribution difference (thin lines).}
\label{fig:7cycle}
\end{figure}

The quantum walk on a line can be converted to an $N$-cycle
by taking a line segment of length $N$, and applying 
periodic boundary conditions.
Clearly, for $t$ greater than $N/2$, when the walk starts to wrap around
on itself, the evolution will be more complicated than a line.
Cycles with odd or even values of $N$ give different results.
For an even cycle, only even (odd) positions
are occupied after an even (odd) number of time steps,
but for an odd cycle, after the first $(N+1)/2$ steps,
both even and odd positions are occupied at the same time. 
We use the same coin operator as for the walk on the line,
given by equation (\ref{eq:genH}).
We observed that the entanglement of the coin, apart from the particular
cases identified by Tregenna et al \cite{tregenna03a}
in which the walk is periodic,
shows no regular pattern, being apparently chaotic. 
We give two examples illustrating this in figure \ref{fig:8+16cycle}.
%%%%%%%% DOUBLE FIGURE %%%%%%%%
Since the entanglement follows the instantaneous state of the system,
it does not tell anything useful about the mixing properties of
the time-averaged distribution.  We also calculated the time-averaged
entanglement (shown in figure \ref{fig:8+16cycle}) and this appears to converge
to a steady value at roughly the same rate as the distribution converges to
its limiting distribution.
This is shown for a seven-cycle in figure \ref{fig:7cycle}.

%----------------------------------------%
\subsection{Cycles with diagonals}
\label{ssec:cyclediags}

\begin{figure}
    \begin{center}
        \resizebox{0.5\columnwidth}{!}{\includegraphics{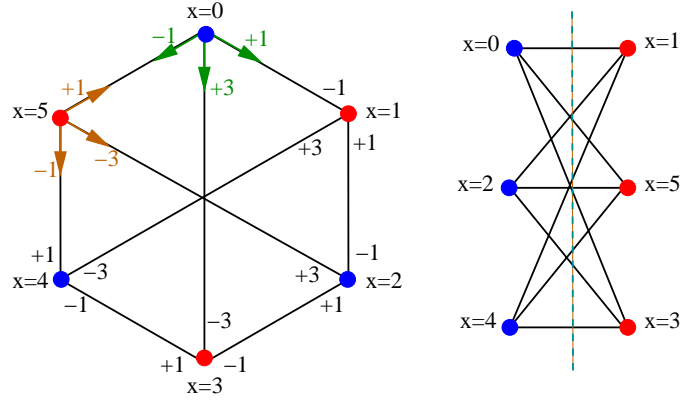}}
    \end{center}
\caption{A six-cycle where the numbers $\pm 1$, and $\pm 3$ label the directions 
	which the particle can take.  This is an example of a complete
	bipartite graph, as redrawn in equivalent form on the right.
	Each blue vertex is connected to every one of the red vertices
	and vice versa; all the edges cross between the two sets
	of vertices (divided by the dashed line).}
\label{fig:6cyclediags}
\end{figure}

Next we consider the case where there are three or more possible directions
that the particle can take. 
As a generalisation of the cycles, we take the case (for $N$ even) where
opposite vertices of the $N$-cycle are joined to give three possible directions
for the particle.
Figure \ref{fig:6cyclediags} explains how this works.
%%%%%%%% FIGURE %%%%%%%%
The third possible path of the particle connects the original position ($x
= 0$) to the opposite position ($x = 3$) of the cycle, and likewise for the
other two opposite pairs.
The edges of the cycle need to be consistently labeled (see
\cite{kendon04a,ambainis04a}), and here we have chosen to
label each end of the edges with either $-1$, 1, $-3$ or 3,
such that adding the vertex and edge label gives
the vertex label at the other end of that edge.
This means that as the particle traverses the edge, the sign of the coin state
must flip, so we adjust the conditional shift operation to act as
\begin{equation}
S\ket{c,v} = \ket{-c,v+c(\text{mod $N$})},
\label{eq:cshiftflip}
\end{equation}
where $v\in \{0\dots(N-1)\}$ is the vertex and $c$ is the coin state,
compare equation (\ref{eq:cshift}).
Although this means we are using more coin states than the degree of the
graph (four instead of three), at any single vertex only three of the four coin
states are actually used, and we pad the coin
operator with zeros (one on the diagonal)
for the unused coin state so it operates correctly
(as a Grover or DFT coin) on the three-dimensional subspace.
More details on how to do this for the general case of a graph with vertices
of various degrees can be found in \cite{kendon04a}. 

\begin{figure}
  \begin{minipage}{0.45\columnwidth}
    \begin{center}
        \resizebox{\columnwidth}{!}{\includegraphics{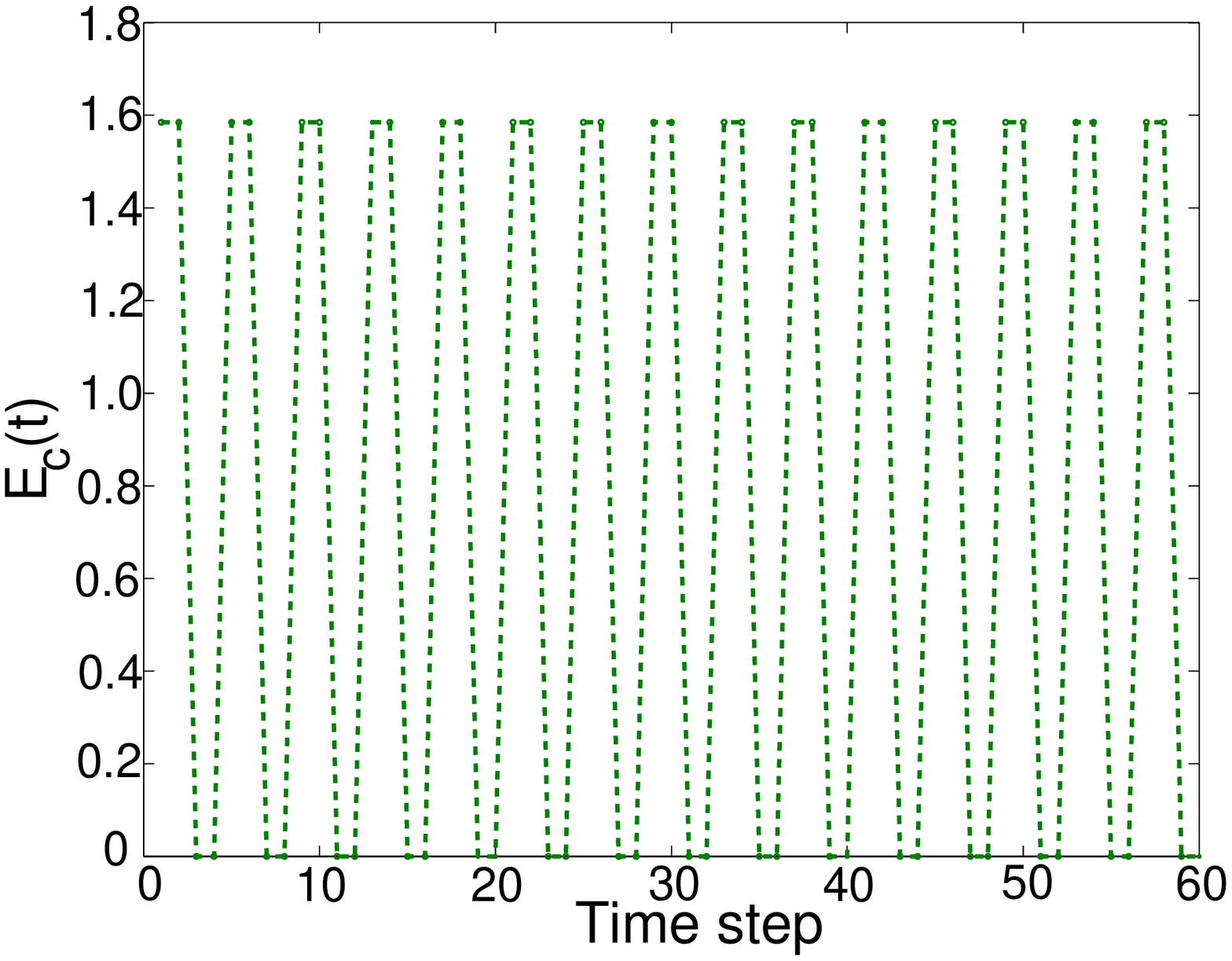}}
    \end{center}
  \end{minipage}
\hfill
  \begin{minipage}{0.48\columnwidth}
    \begin{center}
        \resizebox{\columnwidth}{!}{\includegraphics{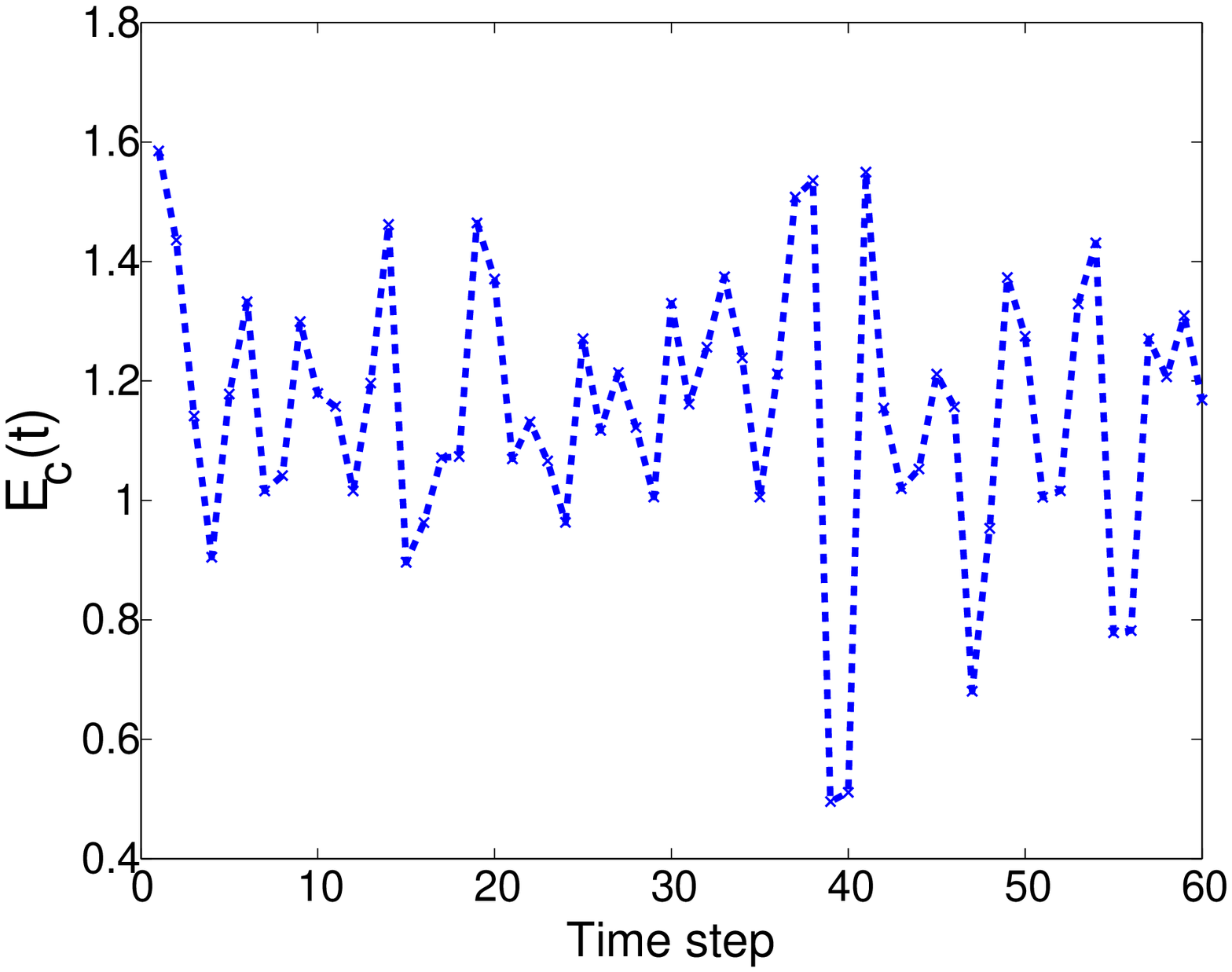}}
    \end{center}
  \end{minipage}
\caption{(Left) Entanglement $E_c(t)$ for the graph of
	figure \ref{fig:6cyclediags} using a
	DFT coin operator and a symmetric initial coin state.
	(Right) Entanglement $E_c(t)$ for 10-cycle with diagonals
	and a DFT coin operator.}
%\textcolor[named]{Red}{\textbf{No source for figure...OK?}}
\label{fig:cdE}
\end{figure}

The evolution of the walk is determined by the coin operator:
we used the $d=3$ Grover and DFT coins,
equations (\ref{eq:grov3}) and (\ref{eq:DFT3}) respectively.
For $N=6$, the entanglement is periodic with period four,
as shown in figure \ref{fig:cdE}
(Left).
%%%%%%%% FIGURE %%%%%%%%
For all other values of $N$ that we tested, the $N$-cycles plus diagonals
showed no regularity, and the entanglement followed a complicated pattern, as 
illustrated in figure \ref{fig:cdE} (Right) for the case of $N=10$.

%----------------------------------------%
\subsection{Complete bipartite graphs}
\label{ssec:bipartite}

The six-cycle with opposite vertices linked is an 
example of a complete bipartite graph. In a bipartite graph
the vertices can be divided into two distinct sets such that
every edge connects between the two sets.
A bipartite graph is complete if, in addition,
each vertex in one set is connected to every vertex in the other set.
These conditions fix the relationship between the
number of vertices and the degree of the graph:
a complete bipartite graph of degree $d$ has $N=2d$ vertices,
we denote it by $K_{d,d}$. 
These graphs can be obtained from the $N$-cycle by adding
edges between every pair of vertices $(i,j)$ in which $i$ is
odd and $j$ is even.
We can label the directions from each
vertex as $\pm1,\pm3,\ldots,\pm(d-1)$, as shown in figure \ref{fig:6cyclediags}
for $d=3$, i.e., $N=6$.
\citeauthor{ahmadi02a} \cite{ahmadi02a} studied the continuous time
quantum walk on similar graphs, looking for instantaneous mixing (i.e.,
a uniform distribution obtained at a particular instant in the time evolution
of the quantum walk).
They found only a small number of examples of instantaneous mixing, for 
regular complete and cyclic graphs with no more than four vertices.
This is in sharp contrast to
classical random walks, which approach a uniform distribution as they
evolve on all well-behaved graphs.
Periodic behaviour is also a property of the instantaneous distributions,
though a slightly less stringent requirement than instantaneous mixing.

We can show analytically that the Grover walk on $K_{d,d}$ is periodic
with period 4, 
by looking at the evolution operator $U=S\cdot (\identity\otimes C_d^{(G)})$
in Fourier space.
With our chosen edge labeling the shift operator acts as
in equation (\ref{eq:cshiftflip}).
Taking a FT on the vertex space only,
\begin{equation}
|v\rangle=\frac{1}{\sqrt{N}}\sum_{k=1}^{N}e^{2\pi ikv/N}|k\rangle,
\end{equation}
shows that the shift acts as
\begin{equation}
S|v,c\rangle=\frac{1}{\sqrt{N}}\sum_{k=1}^{N}e^{2\pi i(v+c)k/N}|k,-c\rangle
\end{equation}
and then
\begin{equation}
S_k|k,c\rangle=e^{2\pi ick/N}|k,-c\rangle,
\end{equation}
meaning that $S$ is block-diagonal in the FT basis. Represent it by 
\begin{equation}
  S_k=
  \left(
    \begin{array}{cccccc}
    0&\omega^k&0&0&\ldots&0\\
    \omega^{-k}&0&0&0&\ldots&0\\
    0&0&0&\omega^{+3k}&\ldots&0\\
    0&0&\omega^{-3k}&0&\ldots&0\\
    \vdots&\vdots&\vdots &\vdots &\ddots &\vdots \\
    0&0&0&0&\ldots&\omega^{-k(d-1)}\\
    \end{array}
  \right),             \label{Sk}
\end{equation}
in which $\omega=e^{2\pi i/N}$, and the last entry is in the
diagonal for the case where $d$ is odd,
otherwise the last block is similar to the others.
In this basis, the ($Nd$-dimensional) evolution operator 
$U=S\cdot(\identity\otimes C_d^{(G)})$ factorises into a set of $N$
matrices $U_k$ each of dimension $d$, with $U_k=S_k\cdot C_d^{(G)}$.
From this one can check explicitly 
that $U_k^4=\identity$, which means that the walk has a period of $4$. 
Alternatively, by looking at the eigenvalues of $U_k$ we can also
verify the periodicity. $U_k$ has eigenvalues $\pm1$ and $e^{i\theta_k}$, with
$\cos\theta_k=\frac{-1}{d}\sum_{c=1,3,\ldots,(d-1)}\cos\frac{2\pi c}{N}$ 
(method used in \cite{ambainis03a}), which reduce to $\pm i$ when $d=N/2$. 

Using numerical simulations with a DFT coin on complete bipartite graphs
we found no examples of periodicity.

%%%%%%%%%%%%%%%%%%%%%%%%%%%%%%%%%%%%%%%%%%%%%%%%
\section{Walks on the ``Glued trees'' graph}
\label{sec:trees}
%----------------------------------------------%

\begin{figure}
  \begin{minipage}{0.45\columnwidth}
    \begin{center}
        \resizebox{0.8\columnwidth}{!}{\includegraphics{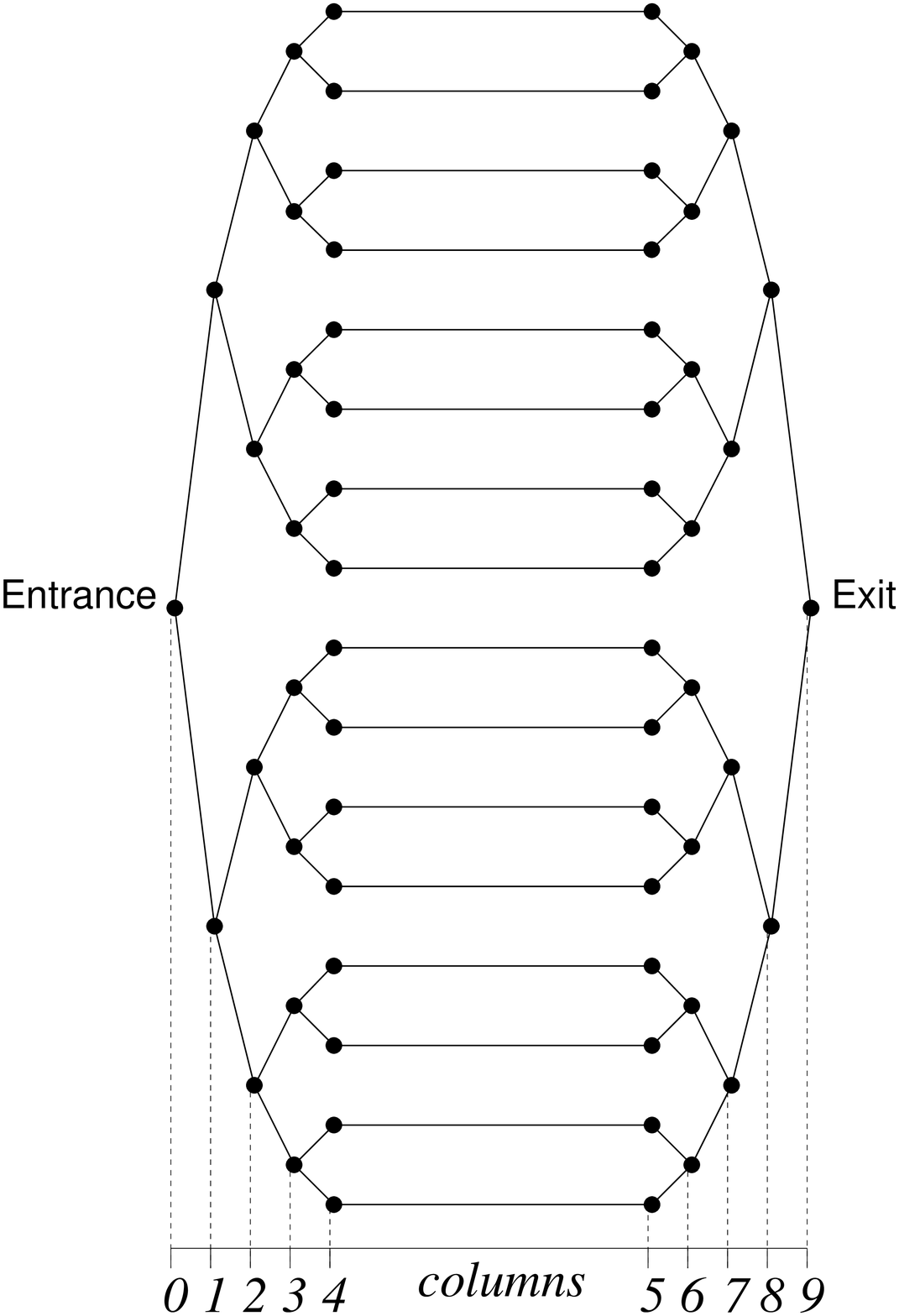}}
    \end{center}
  \end{minipage}
\hfill
  \begin{minipage}{0.45\columnwidth}
    \begin{center}
        \resizebox{0.8\columnwidth}{!}{\includegraphics{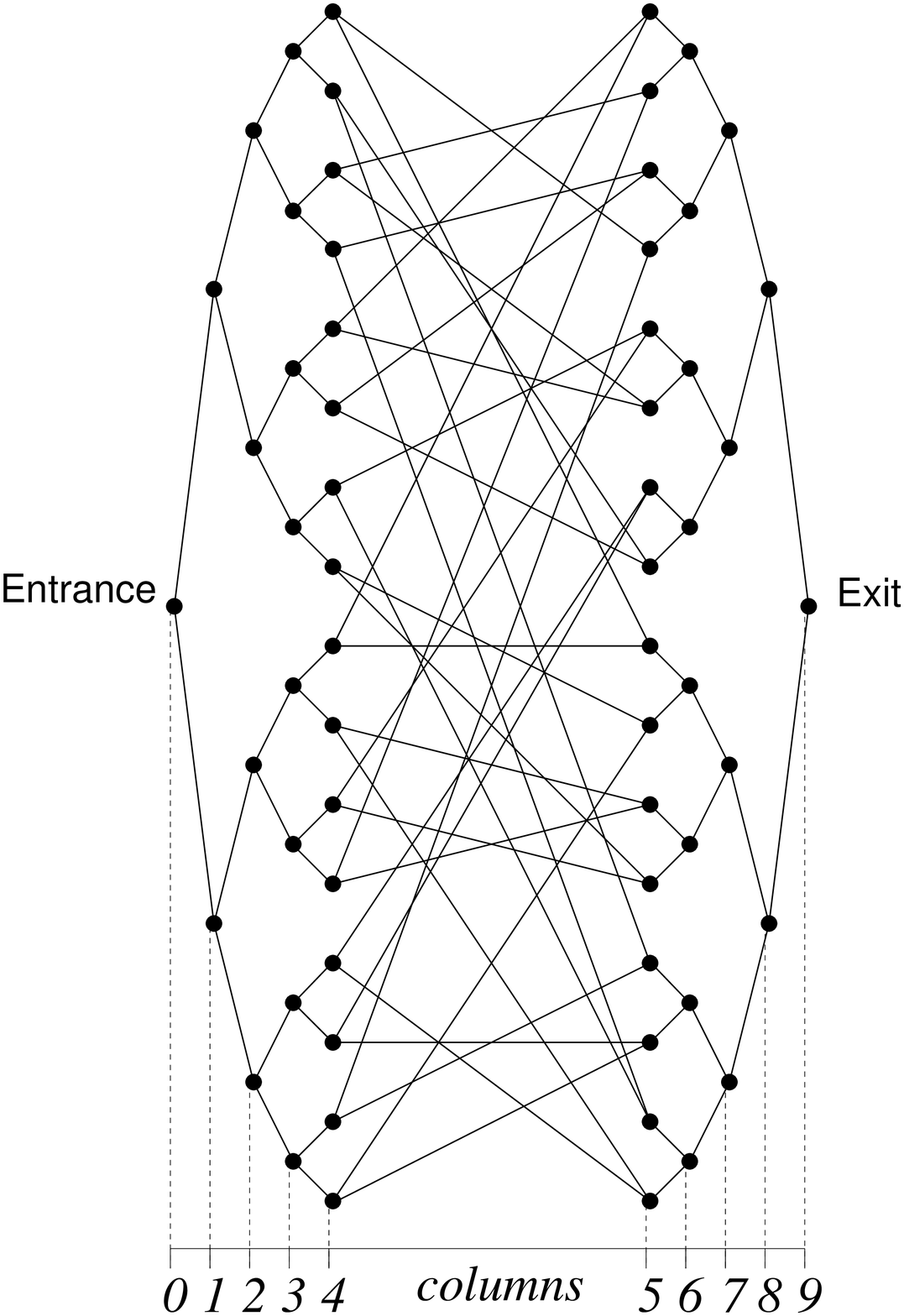}}
    \end{center}
  \end{minipage}
\caption{Left: regularly glued binary tree graph.
         Right:``glued trees'' graph used in the algorithm of
         \cite{childs02a}.  Example shown is for $N=4$, with
         $2N+2=10$ columns 
         and $2(2^{(N+1)}-1)=62$ nodes.
         The gap between columns 4 and 5 is for
         clarity in the figure and is not significant in the algorithm.}
\label{fig:tree4}
\end{figure}

We now turn out attention to the special graph used by
\citeauthor{childs02a} \cite{childs02a} for their algorithm
with an exponential speed up.
An example of this ``glued trees'' graph with tree depth $N=4$ is shown
in figure \ref{fig:tree4} (right).
At the centre, each leaf node has two edges joining it to the leaf
nodes of the other tree so, except for the entrance and exit,
exactly three edges meet at each node.
%%%%%%%% FIGURE %%%%%%%%
The problem is to travel via the edges from node to node
as quickly as possible starting at the entrance and finishing at the exit.
The time taken to reach the exit is an example of a ``hitting time''
(see Kempe \cite{kempe02a} for definitions and an earlier example
of a ``hitting time'' quantum walk problem).
A similar graph with a regular join at the leaves is also shown in
figure \ref{fig:tree4} (left).  This graph is easy to traverse
with a classical algorithm because it is easy to identify the middle
(in this case because the middle nodes have only two edges joining them, but a
regular pattern of two edges per node is also classically tractable).
The randomly glued edges joining the two halves of the graph in 
figure \ref{fig:tree4} (right) disguise the join, and a classical algorithm will
get lost at this point, taking on average exponentially longer to emerge
at the exit.

\begin{figure}
    \begin{center}
        \resizebox{0.6\columnwidth}{!}{\includegraphics{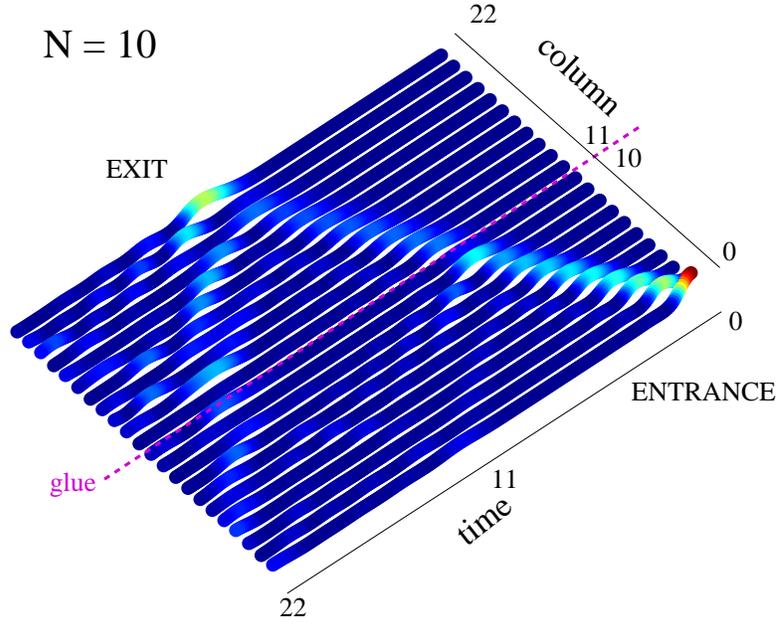}}
    \end{center}
    \caption{Propagation of the continuous time quantum walk
        on a glued trees graph for tree depth $N=10$, run for 44 time steps.}
    \label{fig:contlineprop}
\end{figure}

\citeauthor{childs02a} use a continuous time quantum
walk~\cite{farhi98a} for their algorithm.
The adjacency matrix $\mathbf{A}$ of a graph is an $N\times N$ matrix
with entries $A_{ij}=1$ iff there is an edge joining nodes $i$ and $j$,
all other entries in $\mathbf{A}$ are zero.  For an undirected graph 
(edges can be traversed in either direction, from $i\rightarrow j$
and $j\rightarrow i$), $\mathbf{A}$ is symmetric.  Thus it can be used to
form the Hamiltonian for the quantum walk:
\begin{equation}
i\frac{d}{dt}\braket{x}{\psi(t)} =
        \sum_y^N\bra{x}\mathbf{H}\ket{y}\braket{y}{\psi(t)},
\label{eq:qcon}
\end{equation}
with $\mathbf{H}=\gamma\mathbf{A}$ where $\gamma$ is the transition probability,
and where $x,y$ are nodes in the graph.  The solution may be written
\begin{equation}
\ket{\psi(t)} = 
e^{-i\gamma\mathbf{A}t}\ket{\psi(0)},
\label{eq:qconsol}
\end{equation}
though of course actually calculating it for specific instances of
$\mathbf{A}$ and $\ket{\psi(0)}$ is in general a nontrivial task.
The proof \cite{childs02a} that the quantum walk is exponentially
faster than any classical algorithm involves detailed consideration
of oracles, colourings and simulation of a continuous time quantum
walk on a discrete gate-model quantum computer.
We will not need to discuss these details here.
Figure \ref{fig:contlineprop} shows an example of the propagation of the
continuous time walk through the glued trees graph in terms of the column
positions shown in figure \ref{fig:tree4}.
%%%%%%%% FIGURE %%%%%%%%

\begin{figure}
    \begin{center}
        \resizebox{0.6\columnwidth}{!}{\includegraphics{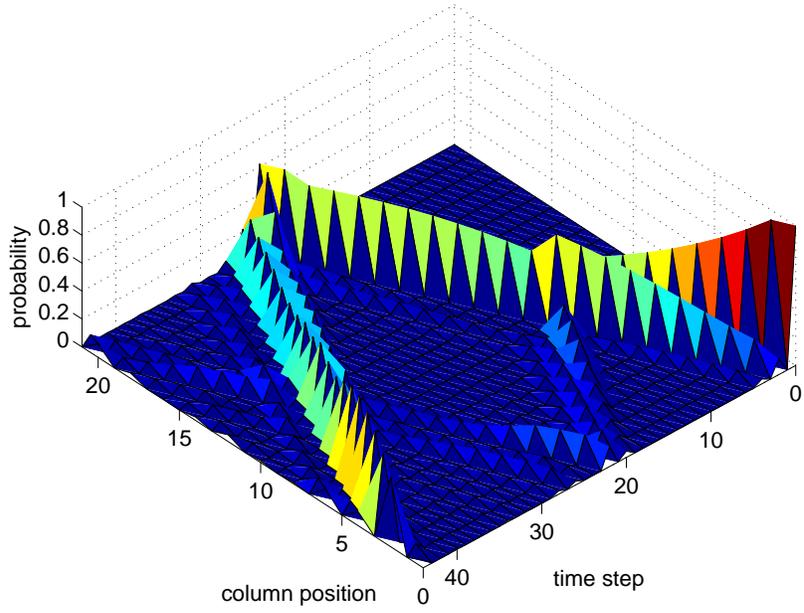}}
    \end{center}
    \caption{Propagation of the discrete quantum walk using the
        Grover coin operator on a glued trees graph
        for tree depth $N=10$, run for 44 time steps.}
        %%(Note to editors: we have an animated version of this graph
        %%for online publication, but provide a 3D static version for
        %%the convenience of the referees.)}
    \label{fig:gtreeprop}
\end{figure}

A discrete time walk can also traverse this graph efficiently
if a three dimensional Grover coin is used
\cite{watrous02a,tregenna03a}.
An example of the propagation using a Grover coin operator
is shown in figure \ref{fig:gtreeprop}.
%%%%%%%% FIGURE %%%%%%%%
As noted by \citeauthor{tregenna03a}~\cite{tregenna03a},
the fast hitting time obtained with a quantum walk is
highly sensitive to the symmetry of the problem:
for quantum walks starting at a node other than the entrance,
the exit becomes exponentially harder to
find and the quantum walk does no better than a classical algorithm.

\begin{figure}
    \begin{center}
        \resizebox{0.5\columnwidth}{!}{\includegraphics{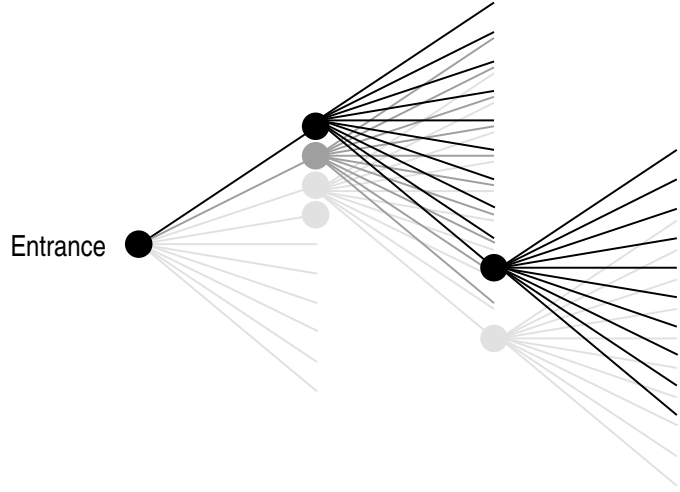}}
    \end{center}
\caption{Detail of a graph with higher branching rate at each node.
        The branching rate in this example is $B=10$, and the degree
        of each node (except the entrance and exit nodes) includes
        the edge from the parent node so is $d=B+1=11$.}
\label{fig:branching}
\end{figure}

\citeauthor{tregenna03a} also noted that
that if a DFT coin operator is used instead of a
Grover coin operator on the ``glued trees''
graph, the quantum walk stays near the starting point 
and does not spread out even as far as a classical random walk.
We consider how this picture changes if we increase the
branching rate of the trees that form the graph.
In other words, we make a similar
graph using a pair of trinary trees (branching rate 3) and quaternary trees
(branching rate 4) and so on for arbitrary branching rate $B$.
This is illustrated in figure \ref{fig:branching}.
%%%%%%%% FIGURE %%%%%%%%

%%%%%%%%%%%%%%%%%%%%%%%%%%%%%%%%%%%%%%%%%%%%%%%%
\subsection{Mapping to a walk on the line}
\label{sec:mapping}
%----------------------------------------------%

Despite the random connections in the centre, the ``glued trees'' graph
with any branching rate is still highly symmetric.
Provided the initial state used at the entrance node respects
the symmetry,
the whole quantum walk process can be mapped to a walk on a line
corresponding to the column positions shown in figure \ref{fig:tree4},
with different biases in the
probabilities for moving right or left at each step. 
The mapping for the continuous time walk is given in
\cite{childs02a} for branching rate $B=2$.
For arbitrary branching rate this generalises to give
a Hamiltonian for column positions
$j$ with non-zero matrix elements
\begin{equation}
\bra{j}H\ket{j+1} = \left\{\begin{array}{ll}
        \sqrt{B}\gamma & 0\le j<N,\,N<j\le 2N \\
        B\gamma        & j=N\\ \end{array} \right.
\end{equation}
and $\bra{j+1}H\ket{j} = \bra{j}H\ket{j+1}$.
We have a choice for the hopping rate $\gamma$.  In order to make a fair
comparison between different branching rates $B$, we take $\gamma=B^{-1/2}$.
This makes the non-zero matrix elements of $H$ unity, except at the glue, 
corresponding to unit hopping rate between column positions in the
mapped-to-line version of the quantum walk, for all choices of $N$ and $B$.

\begin{figure}
    \begin{center}
        \resizebox{0.6\columnwidth}{!}{\includegraphics{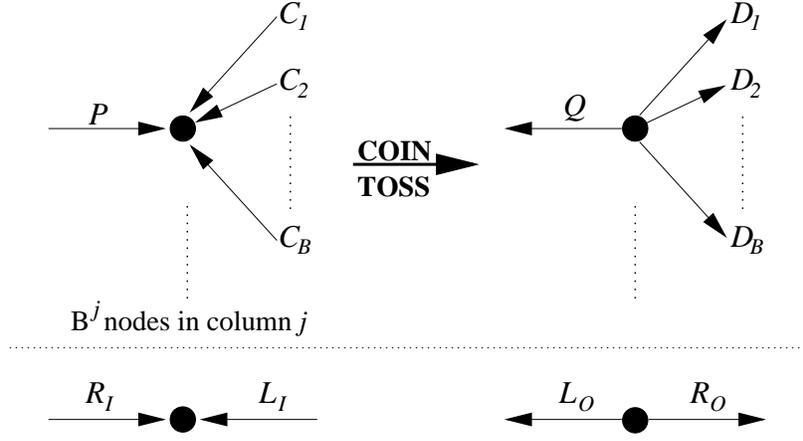}}
    \end{center}
    \caption{Typical node in left hand tree before and after the coin
        operator is applied, showing our notation for
        mapping the walk to a line, $P$, $Q$, $\{C_i\}$, $\{D_i\}$
        $R_I$, $R_O$, $L_I$, $L_O$ are amplitudes.}
    \label{fig:treenode}
\end{figure}

For the discrete time quantum walk the mapping is coin specific,
and only works when the coin operator preserves the symmetry of
the graph so the amplitude of the quantum walk is the same
on all nodes of each column.
To perform the mapping for the Grover coin operator given by
equation (\ref{eq:grovd}), we consider one step of the evolution at a
typical node in the left hand tree.
Our notation is shown in figure \ref{fig:treenode}.
%%%%%%%% FIGURE %%%%%%%%
We assume that $C_1=C_2\dots=C_B=C$ and $D_1=D_2\dots=D_B=D$.
Applying the Grover coin operator gives two relationships between the
incoming and the outgoing amplitudes,
\begin{eqnarray}
   D&=&\frac{2}{B+1}P+\frac{B-1}{B+1}C \nonumber \\
   Q&=&\frac{1-B}{B+1}P+\frac{2B}{B+1}C.
\label{eq:nodeG}
\end{eqnarray}
We then require
$B^j P^2=R^2_I$ and $B^j \sum_{\alpha} C^2_{\alpha}=L^2_I$
for the incoming probabilities on the full tree and on the line,
similarly $B^j Q^{2}=L^2_O$ and $\sum_{\alpha} D^2_{\alpha}=R^2_O$
for the outgoing probabilities.
This gives the relationship between the amplitudes on the full tree
and on the line (independent of choice of coin operator),
\begin{eqnarray}
P=B^{-j/2}R_I && C=B^{-(j+1)/2}L_I \nonumber\\
Q=B^{-j/2}R_O && D=B^{-(j+1)/2}L_O 
\label{eq:mapamp}
\end{eqnarray}
Substituting into equation (\ref{eq:nodeG}) gives the $2\times 2$ operator
for a quantum walk on a line corresponding to the Grover coin operator,
\begin{eqnarray}
  R_O&=&\frac{2\sqrt{B}}{B+1}R_I+\frac{B-1}{B+1}L_I \nonumber\\
  L_O&=&\frac{1-B}{B+1}R_I+\frac{2 \sqrt{B}}{B+1}L_I.
\label{eq:Gmapped}
\end{eqnarray}
The right hand tree is a mirror image of the left hand tree, so for that we just
exchange $L$ and $R$ in the above equations. Unlike the continuous time walk,
there is nothing special for the random edges in the glue, the amplitude for
traversing each edge is determined by the coin operator at the nodes, and this
coin operator is different for each half of the tree.

For the entrance and exit nodes we need to consider the most general choice
in $d$ dimensions that respects the symmetry of the graph.  As explained in
\cite{kendon04a}, although the roots of the tree are only of degree
$d-1=B$, we can pad the extra coin dimension with a piece of the identity
operator so we have $d$ dimensional coins throughout the walk.  This has
the added effect of allowing an extra arbitrary phase, the most general
form of the coin operator at the entrance and exit nodes takes the form
\begin{equation}
C_{\text{\textsc{end}}} = \left( \begin{array}{cc}
        e^{i\phi}C^{(B)} & \mathbf{0} \\
        \mathbf{0} & 1 \\ \end{array} \right),
\label{eq:cgrovB}
\end{equation}
where $C^{(B)}$ is a suitably symmetric coin operator of dimension $B=d-1$,
with an extra arbitrary phase $\phi$, and the boldface zeros fill in the
row and column to make $C_{\text{\textsc{end}}}$ have dimension $d$.
We then map $C_{\text{\textsc{end}}}$ to a walk on the line, which
gives a reflection with a phase shift
(that may be chosen independently at the entrance and the exit),
\begin{eqnarray}
  R_O&=&e^{i\phi_1}L_I \text{\null\hspace{1em} \textsc{Entrance}}\nonumber\\
  L_O&=&e^{i\phi_2}R_I \text{\null\hspace{1em} \textsc{Exit}}.
\end{eqnarray}

This mapped-to-the-line operator already tells us how the Grover coin
operator will behave in the limit of large branching rate.  
Equation~(\ref{eq:Gmapped}) becomes $-i\sigma_y$ as $B\rightarrow\infty$,
for which the walk on a line simply oscillates between the initial and 
neighbouring nodes, making no further progress along the line.
Thus we expect the probability of reaching the exit to fall towards
zero as $B\rightarrow\infty$.

We can relate the entanglement between the coin and the position on
the line to the entanglement between the coin and the position
in the walk on the full graph.
To calculate the entanglement, we calculate the entropy of the reduced
density matrix for the coin (by tracing over the position)
and then obtain the eigenvalues of this density matrix, from which
the entropy is calculated according to equation (\ref{eq:entdef}).
For a two-dimensional matrix the eigenvalues are the solution of
a quadratic equation, for a $d$-dimensional coin the eigenvalues
are the solution of a polynomial of order $d$.
By calculating the reduced density matrices for the full walk and the 
mapped-to-line versions, and comparing the polynomials we find,
for example, for the Grover coin case, the polynomials are
related by
\begin{equation}
f^{(B)}_{\textsc{full}}(x) = (-x)^{B-1}(x^2 - (a_B + B d_B)x + B(a_Bd_B - c^2_B)) =
(-x)^{B-1}f^{(B)}_{\textsc{line}}(x),
\end{equation}
where $a_B$, $c_B$, $d_B$ are time dependent coefficients.
The roots of this polynomial give the eigenvalues and thence the entropy
of the reduced density matrix, which measures the entanglement.
Thus we see that the entanglement is the same between the coin and position
in the mapped-to-line walk as in the corresponding full walk.
We studied the entanglement for the walk on the glued trees graph, but it
did not provide useful indications of the progress of the walk, so we
do not present any of these results here.

%%%%%%%%%%%%%%%%%%%%%%%%%%%%%%%%%%%%%%%%%%%%%%%%
\subsection{Variation with tree depth $N$}
\label{sec:depth}
%----------------------------------------------%

We first restrict our attention to the original problem with branching
rate $B=2$, and consider what happens when the tree depth $N$ is varied.
The quantum walker does not arrive the exit with certainty after
a short number of time steps.  However, the probability of finding it
at the exit node shows a clear peak after roughly the number of steps
it takes to walk deterministically from entrance to exit by the shortest
route.  If this peak is large enough (polynomial in $N$ and $B$),
one can simply repeat the quantum walk a polynomial number of times to increase
the success probability to near certainty (probability amplification).

%%%%%%%% FIGURE %%%%%%%%
\begin{figure}
    \begin{center}
        \resizebox{0.6\columnwidth}{!}{\includegraphics{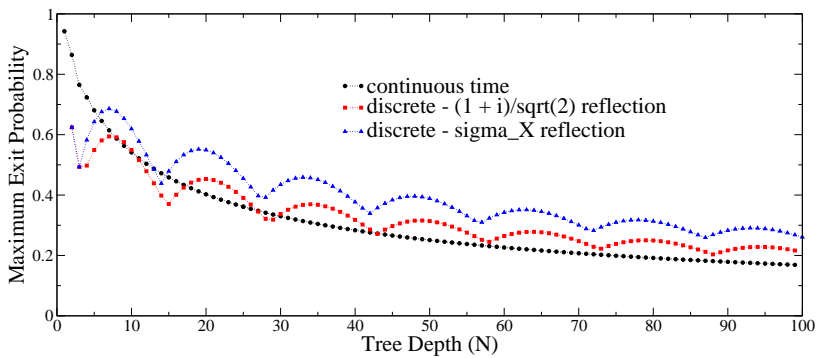}}
    \end{center}
    \caption{Variation of exit probability with tree depth on the ``glued
        trees'' graph with branching rate $B=2$ for the continuous
        time quantum walk (black) and the discrete quantum walk with a
        Grover coin using two different reflection phases as detailed
        in the key (red and blue).}
    \label{fig:treedepth}
\end{figure}
We studied both continuous time and discrete time quantum walks
using simulations with the dynamics mapped onto a walk on the line.
For the continuous time walk the results are straightforward, given by the
black line in figure \ref{fig:treedepth}.  This is well-fit for large $N$
by $P(\text{\textsc{exit}})\sim N^{-2/3}$, as predicted by
\citeauthor{childs02a} \cite{childs02a},
who derive the scaling of the Green's function to be $\sim N^{-1/3}$
at the peak.

For the discrete time walk, as noted in the previous section,
we have a choice about what to do at the entrance
and exit where the nodes are of degree two rather than three.
We tested two examples of choices of phase shifts, equation (\ref{eq:cgrovB}),
using the same phase at both entrance and exit.
The 2D version of the Grover coin operator is $\sigma_x$,
which produces no phase shift when mapped to the line.
Another possible symmetric 2D coin operator is
\begin{equation}
\mathbf{C}_2^{(\text{S})}=\frac{1}{\sqrt{2}}\left( \begin{array}{rr}
        1 & i \\
        i & 1 \\
        \end{array} \right).
\label{eq:2Dsym}
\end{equation}
Mapped to the line, this produces a phase shift of $(1+i)/\sqrt{2}=e^{i\pi/4}$.
The different phases give different results due to the different interference
effects.  Results for both are also shown in figure \ref{fig:treedepth}.
They produce a slightly higher exit probability than the continuous
time walk for most values of $N$, but follow the same scaling
of $P(\text{\textsc{exit}})\sim N^{-2/3}$ (numerical data up to
$N=10000$ not shown in figure \ref{fig:treedepth} was used for the fitting).
This corresponds with the analytic solution for a simple walk on a line
found by \citeauthor{ambainis01a} \cite{ambainis01a},
which gives the scaling of the peak amplitude as $\sim N^{-1/3}$.
The mapped-to-the line quantum walk is not exactly the same as the simple walk
on an infinite line, but the scaling of the peak should be similar, 
up to the first reflection at the exit.
This scaling, being polynomial in $N$, does not affect the exponential
speed up of the algorithm.  We can apply probability amplification
efficiently in polynomial time and still beat the exponentially
slow classical algorithm.
However, the difference between the two different choices of
2D coin operators is instructive, suggesting that in other algorithms an
appropriate choice might significantly improve the success probability.

%%%%%%%%%%%%%%%%%%%%%%%%%%%%%%%%%%%%%%%%%%%%%%%%
\subsection{Variation with branching rate}
\label{sec:branch}
%----------------------------------------------%

We now consider how the exit probability varies if the branching rate of the
two trees forming the graph is varied.
For the continuous time walk, \citeauthor{childs02a} \cite{childs02a}
provide most of the calculations,
they give the transmission amplitude through a ``defect'' (i.e.~the glue) of 
\begin{equation}
\mathcal{T}(\kappa) = \frac{2i\sqrt{B}\sin\kappa}{(B-1)\cos\kappa + (B+1)\sin\kappa},
\end{equation}
(in our notation) where $0<\kappa<\pi$ is the momentum of the quantum walker.
This has a broad peak at $\kappa=\pi/2$, where the transmission probability
$|\mathcal{T}(\pi/2)|^2=4B/(B+1)^2$.  Thus for large $B$ we
expect the success probability to scale as $1/B$, and we observed this
numerically.

\begin{figure}
  \begin{minipage}{0.6\columnwidth}
    \begin{center}
        %\resizebox{\columnwidth}{!}{\includegraphics{eps/grvPpeak50.eps}}
	\resizebox{\columnwidth}{!}{\rotatebox{-90}{\includegraphics{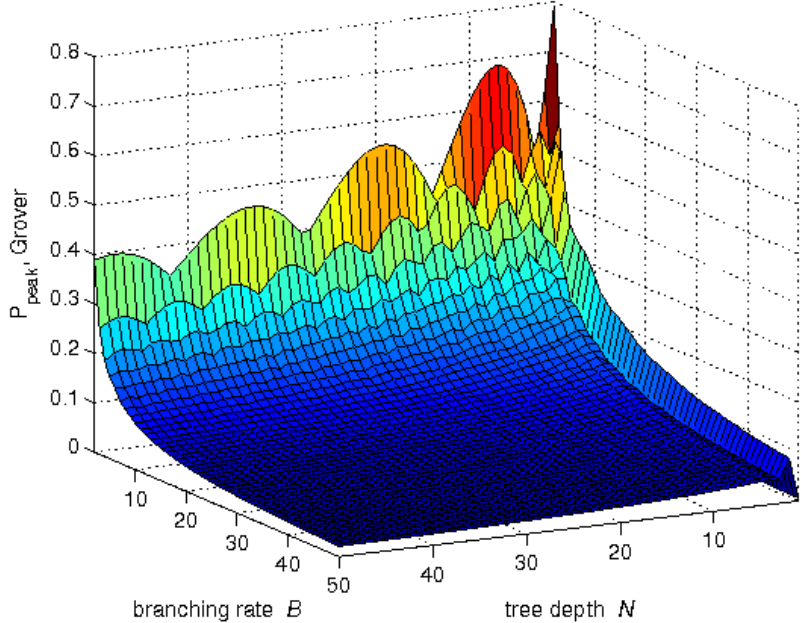}}}
    \end{center}
  \end{minipage}
\caption{Variation of the first peak in the exit probability
        for branching rates up to $B=50$, and tree depths up to $N=50$
        using a Grover coin operator.}
\label{fig:treebranchgrov}
\end{figure}

For the discrete time walk the results are shown in
figure \ref{fig:treebranchgrov}.
%%%%%%%% FIGURE %%%%%%%%
The peak exit probability closely follows that of the continuous time walk
scaling as $1/B$.  Thus both continuous time and discrete Grover coined
walks still beat a classical algorithm, which takes time
proportional to the number of nodes, i.e.~$T\sim B^N$, to find the exit
with better than exponentially small exit probability.
Combining the scalings for depth and branching rate,
we have these two versions of quantum walks finding the exit
in time linear in $N$ with probability proportional to $1/(BN^{2/3})$.

%%%%%%%%%%%%%%%%%%%%%%%%%%%%%%%%%%%%%%%%%%%%%%%%
\subsection{DFT coin operator on glued trees}
\label{sec:dftgtrees}
%----------------------------------------------%

\begin{figure}
    \begin{center}
        \resizebox{0.6\columnwidth}{!}{\includegraphics{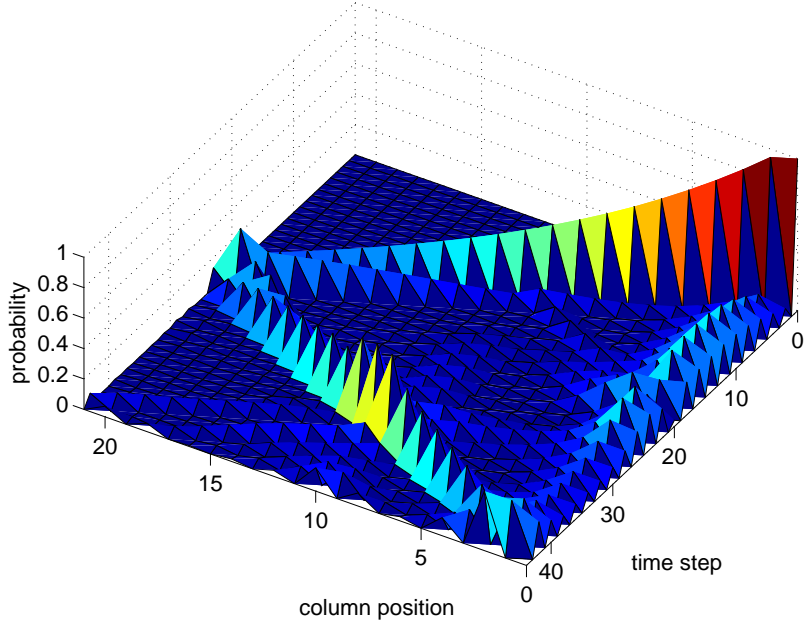}}
    \end{center}
    \caption{Propagation of the discrete quantum walk using the
        DFT coin operator on a glued trees graph for tree depth
        $N=10$ and branching rate $B=10$, run for 44 time steps.}
    \label{fig:dtreeprop}
\end{figure}

We now consider the quantum walk with a DFT coin operator on
the glued trees graph. 
The DFT coin operator 
does not have the right symmetry to map to a walk on the line
for randomly applied coin labels, but if we apply a regular
coin labeling that singles out the ``root'' direction to have coin label zero
the mapping to the line can be carried out successfully.
This is ``cheating'' as far as application to traversing the
graph is concerned, since such a labeling would enable a classical
algorithm to deterministically find the exit.
Nonetheless, the branching trees that make up this graph appear
widely in physics and computer science, so other applications that could
use this regular labeling may arise.
Note that the coin labeling in \cite{tregenna03a}
was consistent in that each end of each edge had the same label, but 
therefore could not always label the root direction as zero, and in fact
made random label choices once the consistency condition was satisfied.
Hence they did not find any advantage to using  DFT coin over classical.
With the regular labeling, the mapping works because the phase
factors all neatly cancel out, as we now show.
We start by assuming $C_1=C_2\dots=C_B=C$, and show that after
applying the DFT coin operator as in equation (\ref{eq:DFTd}) for $d=B+1$ we have
$D_1=D_2\dots=D_B=D$.  Thus if we start in a symmetric state,
this will be preserved under the DFT coin operator.
Using the notation in figure \ref{fig:treenode},
\begin{eqnarray}
   Q  &=&\frac{1}{\sqrt{B+1}}\left\{P+BC\right\} \nonumber \\
   D_1&=&\frac{1}{\sqrt{B+1}}\left\{P+(\omega+\omega^2+\dots+\omega^B)C\right\} 
\nonumber\\
   D_2&=&\frac{1}{\sqrt{B+1}}\left\{P+(\omega^2+\omega^4+\dots+\omega^{2B})C\right\} \nonumber\\
   &\vdots& \nonumber\\
   D_B&=&\frac{1}{\sqrt{B+1}}\left\{P+(\omega^B+\omega^{2B}+\dots+\omega^{B^2})C
\right\} 
\label{eq:nodeDFT}
\end{eqnarray}
where $\omega=e^{2\pi i/d}$ is the complex $d$'th root of unity.
Each of the sums of powers of $\omega$ in the expressions for $D_1$ to $D_B$
evaluates to $-1$, hence we have $D_1=D_2\dots=D_B=D$ as claimed.
Substituting for $P$, $Q$, $C$ and $D$ from equation (\ref{eq:mapamp})
then gives
\begin{eqnarray}
  R_O&=&\sqrt{\frac{B}{B+1}}R_I-\frac{1}{\sqrt{B+1}}L_I \nonumber\\
  L_O&=&\frac{1}{\sqrt{B+1}}R_I+\sqrt{\frac{B}{B+1}}L_I 
\label{eq:Dmapped}
\end{eqnarray}
for the $2\times 2$ operator equivalent to the DFT coin operator.
For $B\rightarrow\infty$, equation (\ref{eq:Dmapped}) becomes the identity, which
corresponds to a walk that steps deterministically along the line,
reaching the exit in the shortest possible time.
For intermediate values of $B$, the DFT coined walk partly steps across the
graph, and partly remains close to the entrance: an example is shown in
figure \ref{fig:dtreeprop}
for a ``glued trees'' graph with branching rate $B=10$ and depth $N=10$.
%%%%%%%% FIGURE %%%%%%%%

Again we look at the first peak in
the exit probability, which occurs after time $\sim(2N+2)$.
%%%%%%%% FIGURE %%%%%%%%
\begin{figure}
  \begin{minipage}{0.6\columnwidth}
    \begin{center}
        %\resizebox{\columnwidth}{!}{\includegraphics{eps/dftPpeak50.eps}}
	\resizebox{\columnwidth}{!}{\rotatebox{-90}{\includegraphics{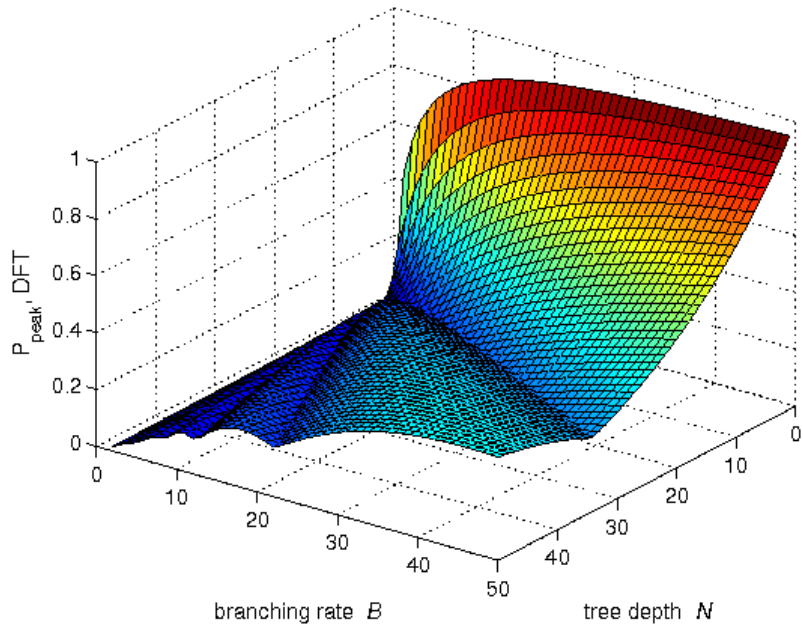}}}
    \end{center}
  \end{minipage}
\caption{Variation of the first peak in the exit probability
        for branching rates up to $B=50$, and tree depths up to $N=50$
        using a DFT coin operator.}
\label{fig:treebranchdft}
\end{figure}
The variation of this first peak in the probability
with branching rate and tree depth for $N,B<50$ is plotted
in figure \ref{fig:treebranchdft}.
For $B=N$, there is a wide peak; on the $B>N$ side an initial peak rises
towards unity, while on the $B<N$ side the probability tails off in 
a series of smaller arches. 
For $B=2$, the quantum walk with this coin operator
has a noticeably lower probability of finding the exit after linear time
than the Grover coin or the continuous time quantum walks.
Nevertheless, there is still a peak in the probability of finding the
walker at the exit, and our numerical results show this
peak scales as $N^{-2/3}$, the same as the Grover coined walk
and the continuous time walk, but with a constant prefactor 
$\sim 30$ times smaller.

We can understand this prefactor qualitatively as follows.
The mapped-to-the-line version of the coined quantum walk is
essentially a generalised Hadamard coin walk on finite line segments
\cite{bach02a}.
Such walks on the infinite line all have the typical double-peaked shape,
with the ratio of the left and right peak heights determined by
the coin biases and the initial state.
The mapped-to-the-line versions of the Grover and regular DFT
coins have the minus sign in opposite positions in the off-diagonal,
equations (\ref{eq:Gmapped}) and (\ref{eq:Dmapped}).  Combined with the rightward
moving initial state, for the first half of the walk
the largest peak is thus on the leading(trailing) edge for
the Grover(DFT) coin operator.  Moreover, the trailing edge DFT peak
is larger than the leading edge Grover peak.
For the second half of the walk the coin 
operators are transposed, and the trailing(leading) edge is largest for
the Grover(DFT) coin operator, but this is combined with different
``initial'' states as the quantum walk crosses the glue from the first
half of the walk, so the two effects don't just cancel out.
The cross-over point beyond which the regular DFT coin performs
best is roughly $B>10$ for any tree depth.

%%%%%%%%%%%%%%%%%%%%%%%%%%%%%%%%%%%%%%%%%%%%%%%%
\section{Summary and Discussion}
\label{sec:summary}
%----------------------------------------------%

Even for the simple case of a coined quantum walk on the line
we find that the entanglement between the coin and the particle position
shows surprisingly rich behaviour.  The entanglement oscillates
around an asymptotic value that is determined by the bias in the coin
operator, and the rate of convergence depends the symmetry in the
distribution of the quantum walk, which in turn is determined by the initial
state of the coin (assuming the quantum walk starts at the origin).
More symmetric distributions (about the origin) converge faster than
asymmetric distributions in which there is a bias favouring finding
the particle in the positive or the negative half line.

For lattices in two spatial dimensions, similar rich patterns of
behaviour are observed, the rate of convergence of the entanglement
correlates with the extent to which the quantum walk spreads out over
the grid, with fast convergence indicating good spreading behaviour,
and slow convergence for distributions that bunch around one point,
such as the origin.
We find that for lattices in two spatial dimensions but of higher
degree $d>4$, a DFT coin operator is better at spreading
over the graph than a Grover coin operator.
This can be explained by the bias in the Grover coin operator: above
degree $d=4$, the Grover coin operator favours returning along the path
it arrived from rather than picking a new direction, thus tending to keep
the walk near where it started.  The DFT coin operator is unbiased for
any degree, so the spreading depends only on the relative phases, which can be controlled by the coin labeling.  

For finite graphs, there are a few periodic examples, corresponding
to those reported in \cite{tregenna03a}, otherwise the entanglement is
random even for very simple cases, like the seven-cycle.
A quantum walk on a complete bipartite graphs using a Grover coin
operator is always periodic with period four, and we have shown this
analytically.  For the DFT coin, the quantum walk on complete bipartite
graphs was not periodic for any cases we tested numerically.

For the glued trees graph and its generalisation to higher branching rates,
the first peak in the probability for finding the quantum walker at the
exit node occurs in linear time and scales 
with tree depth as $N^{-2/3}$ for both Grover and DFT coin
operators, matching that found for the continuous time walk by
\citeauthor{childs02a} \cite{childs02a}.  This
is in contrast with classical algorithms, which require exponential time
to find the exit with better than exponentially small probability.
However, for increased branching rate,
while the continuous time walk and the Grover coined discrete time walk
both scale as $B^{-1}$, a  DFT coin with regular coin labels
has a success probability approaching unity for $B>N>>1$,
and beats the continuous time and Grover coined walks for $B\gtrsim 10$.
A regular coin labeling is not permitted in the algorithmic context of
\cite{childs02a}, but branching trees occur in many other contexts, some
of which may permit a regular coin labeling such that this DFT coin
property could prove useful.

Another observation from this work is that since the coin needs to have
as many dimensions as the maximum degree of the graph, it
allows a relative phase to be introduced at vertices of less than the maximum
degree, for example, at the entrance and exit of the ``glued trees'' graph,
see figure \ref{fig:treedepth}.
The quantum walk may exhibit different properties depending on the choice of
relative phase, which provides extra options for tuning the quantum walk.

Overall, the discrete quantum walk offers many options for tuning the
properties of the quantum coin operator to fit the problem to be solved.
These possibilities are not available in the continuous
time quantum walk, unless one adds extra degrees of freedom
\cite{childs04a} (whereupon it becomes exactly equivalent).
Further work could investigate how to optimise the uniformity of
spreading over these higher dimensional lattices, for example, by
applying small amounts of decoherence (see \cite{tregenna03a}) and by
adjusting the coin operator: we do not know whether the DFT coin operator
is the optimal choice for a general $d$-dimensional graph for $d>4$.

%---------------------%
\begin{acknowledgments}
%---------------------%

We thank many people for interesting discussions of quantum walks:
among them,
Andris Ambainis,
Jamie Batuwantudawe (who contributed to preliminary work on the scaling
with tree depth),
Andrew Childs,
Richard Cleve,
Ed Farhi,
Peter H\o{}yer,
Cris Moore,
Eugenio Rold\'{a}n,
Alex Russell,
Barry Sanders,
John Sipe,
Susan Stepney (who first asked the question about high fan out trees),
Ben Tregenna,
and
John Watrous
stimulated our thinking for the work in this paper.
This work was funded in part by
CAPES Ministry of Education of Brazil,
the European Union and
the UK Engineering and Physical Sciences Research Council.

We learned with much sadness of the untimely death of author Ivens Carneiro
in a road accident in April 2006.

Note added April 2008: we thank Annette Gattner and Mostafa Annabestani
for independently pointing out some errors in our published
figure 2 (left) and related results in section \ref{ssec:entropylimit}.

%---------------------%
\end{acknowledgments}
%---------------------%

%%%%%%%%%%%%%%%%%%%%%%%%%%%%%%%%%%%%%%%%%%%%%%%%%%%%%%%%%%%%%%%%%%%%%%%%%%%

%-------------------------------------%
%%%% construct refs with natbib... %%%%
% single spacing for refs while in draft format: %
\renewcommand{\baselinestretch}{1.0}\small\normalsize
\bibliography{../bibs/qrw,../bibs/qit}

\begin{thebibliography}{32}
\expandafter\ifx\csname natexlab\endcsname\relax\def\natexlab#1{#1}\fi
\expandafter\ifx\csname bibnamefont\endcsname\relax
  \def\bibnamefont#1{#1}\fi
\expandafter\ifx\csname bibfnamefont\endcsname\relax
  \def\bibfnamefont#1{#1}\fi
\expandafter\ifx\csname citenamefont\endcsname\relax
  \def\citenamefont#1{#1}\fi
\expandafter\ifx\csname url\endcsname\relax
  \def\url#1{\texttt{#1}}\fi
\expandafter\ifx\csname urlprefix\endcsname\relax\def\urlprefix{URL }\fi
\providecommand{\bibinfo}[2]{#2}
\providecommand{\eprint}[2][]{\url{#2}}

\bibitem[{\citenamefont{Childs et~al.}(2003)\citenamefont{Childs, Cleve,
  Deotto, Farhi, Gutmann, and Spielman}}]{childs02a}
\bibinfo{author}{\bibfnamefont{A.~M.} \bibnamefont{Childs}},
  \bibinfo{author}{\bibfnamefont{R.}~\bibnamefont{Cleve}},
  \bibinfo{author}{\bibfnamefont{E.}~\bibnamefont{Deotto}},
  \bibinfo{author}{\bibfnamefont{E.}~\bibnamefont{Farhi}},
  \bibinfo{author}{\bibfnamefont{S.}~\bibnamefont{Gutmann}}, \bibnamefont{and}
  \bibinfo{author}{\bibfnamefont{D.~A.} \bibnamefont{Spielman}}, in
  \emph{\bibinfo{booktitle}{Proc.~35th Annual ACM STOC}}
  (\bibinfo{publisher}{ACM, NY}, \bibinfo{year}{2003}), pp.
  \bibinfo{pages}{59--68}, \bibinfo{note}{arXiv:quant-ph/0209131},
  \eprint{ArXiv:quant-ph/0209131}.

\bibitem[{\citenamefont{Shenvi et~al.}(2003)\citenamefont{Shenvi, Kempe, and
  {Birgitta Whaley}}}]{shenvi02a}
\bibinfo{author}{\bibfnamefont{N.}~\bibnamefont{Shenvi}},
  \bibinfo{author}{\bibfnamefont{J.}~\bibnamefont{Kempe}}, \bibnamefont{and}
  \bibinfo{author}{\bibfnamefont{K.}~\bibnamefont{{Birgitta Whaley}}},
  \bibinfo{journal}{Phys.~Rev.~A} \textbf{\bibinfo{volume}{67}},
  \bibinfo{pages}{052307} (\bibinfo{year}{2003}),
  \bibinfo{note}{arXiv:quant-ph/0210064}, \eprint{ArXiv:quant-ph/0210064}.

\bibitem[{\citenamefont{Childs and Eisenberg}(2005)}]{childs03b}
\bibinfo{author}{\bibfnamefont{A.}~\bibnamefont{Childs}} \bibnamefont{and}
  \bibinfo{author}{\bibfnamefont{J.~M.} \bibnamefont{Eisenberg}},
  \bibinfo{journal}{Quantum Information and Computation}
  \textbf{\bibinfo{volume}{5}}, \bibinfo{pages}{593} (\bibinfo{year}{2005}),
  \bibinfo{note}{arXiv:quant-ph/0311038}, \eprint{ArXiv:quant-ph/0311038}.

\bibitem[{\citenamefont{Magniez et~al.}(2005)\citenamefont{Magniez, Santha, and
  Szegedy}}]{magniez05a}
\bibinfo{author}{\bibfnamefont{F.}~\bibnamefont{Magniez}},
  \bibinfo{author}{\bibfnamefont{M.}~\bibnamefont{Santha}}, \bibnamefont{and}
  \bibinfo{author}{\bibfnamefont{M.}~\bibnamefont{Szegedy}}, in
  \emph{\bibinfo{booktitle}{Proceedings of 16th ACM-SIAM Symposium on Discrete
  Algorithms}} (\bibinfo{publisher}{Society for Industrial and Applied
  Mathematics, Philadelphia, PA, USA}, \bibinfo{year}{2005}), pp.
  \bibinfo{pages}{1109--1117}.

\bibitem[{\citenamefont{Ambainis}(2004)}]{ambainis03a}
\bibinfo{author}{\bibfnamefont{A.}~\bibnamefont{Ambainis}}, in
  \emph{\bibinfo{booktitle}{45th Annual IEEE Symposium on Foundations of
  Computer Science, Oct 17-19, 2004}} (\bibinfo{publisher}{IEEE Computer
  Society Press, Los Alamitos, CA}, \bibinfo{year}{2004}), pp.
  \bibinfo{pages}{22--31}, \eprint{quant-ph/0311001}.

\bibitem[{\citenamefont{Kempe}(2003{\natexlab{a}})}]{kempe03a}
\bibinfo{author}{\bibfnamefont{J.}~\bibnamefont{Kempe}},
  \bibinfo{journal}{Contemp.~Phys.} \textbf{\bibinfo{volume}{44}},
  \bibinfo{pages}{302} (\bibinfo{year}{2003}{\natexlab{a}}),
  \eprint{quant-ph/0303081}.

\bibitem[{\citenamefont{Ambainis}(2003)}]{ambainis04a}
\bibinfo{author}{\bibfnamefont{A.}~\bibnamefont{Ambainis}},
  \bibinfo{journal}{Intl.~J.~ Quantum Information}
  \textbf{\bibinfo{volume}{1}}, \bibinfo{pages}{507} (\bibinfo{year}{2003}),
  \bibinfo{note}{arXiv:quant-ph/0403120}, \eprint{ArXiv:quant-ph/0403120}.

\bibitem[{\citenamefont{Jerrum et~al.}(2001)\citenamefont{Jerrum, Sinclair, and
  Vigoda}}]{jerrum01a}
\bibinfo{author}{\bibfnamefont{M.}~\bibnamefont{Jerrum}},
  \bibinfo{author}{\bibfnamefont{A.}~\bibnamefont{Sinclair}}, \bibnamefont{and}
  \bibinfo{author}{\bibfnamefont{E.}~\bibnamefont{Vigoda}}, in
  \emph{\bibinfo{booktitle}{Proc.~33rd Annual ACM STOC}}
  (\bibinfo{publisher}{ACM, NY}, \bibinfo{year}{2001}), pp.
  \bibinfo{pages}{712--721}.

\bibitem[{\citenamefont{Sch{\"o}ning}(1999)}]{schoning99a}
\bibinfo{author}{\bibfnamefont{U.}~\bibnamefont{Sch{\"o}ning}}, in
  \emph{\bibinfo{booktitle}{40th Annual Symposium on FOCS}}
  (\bibinfo{publisher}{IEEE, Los Alamitos, CA}, \bibinfo{year}{1999}), pp.
  \bibinfo{pages}{17--19}.

\bibitem[{\citenamefont{Dyer et~al.}(1991)\citenamefont{Dyer, Frieze, and
  Kannan}}]{dyer91a}
\bibinfo{author}{\bibfnamefont{M.}~\bibnamefont{Dyer}},
  \bibinfo{author}{\bibfnamefont{A.}~\bibnamefont{Frieze}}, \bibnamefont{and}
  \bibinfo{author}{\bibfnamefont{R.}~\bibnamefont{Kannan}},
  \bibinfo{journal}{J.~of the {ACM}} \textbf{\bibinfo{volume}{38}},
  \bibinfo{pages}{1} (\bibinfo{year}{1991}).

\bibitem[{\citenamefont{Motwani and Raghavan}(1995)}]{motwani95}
\bibinfo{author}{\bibfnamefont{R.}~\bibnamefont{Motwani}} \bibnamefont{and}
  \bibinfo{author}{\bibfnamefont{P.}~\bibnamefont{Raghavan}},
  \emph{\bibinfo{title}{Randomized Algorithms}} (\bibinfo{publisher}{Cambridge
  University Press}, \bibinfo{address}{Cambridge, UK}, \bibinfo{year}{1995}).

\bibitem[{\citenamefont{{Aharonov, Y} et~al.}(1992)\citenamefont{{Aharonov, Y},
  Davidovich, and Zagury}}]{aharonov92a}
\bibinfo{author}{\bibnamefont{{Aharonov, Y}}},
  \bibinfo{author}{\bibfnamefont{L.}~\bibnamefont{Davidovich}},
  \bibnamefont{and} \bibinfo{author}{\bibfnamefont{N.}~\bibnamefont{Zagury}},
  \bibinfo{journal}{Phys.~Rev.~A} \textbf{\bibinfo{volume}{48}},
  \bibinfo{pages}{1687} (\bibinfo{year}{1992}).

\bibitem[{\citenamefont{Watrous}(2001)}]{watrous98a}
\bibinfo{author}{\bibfnamefont{J.}~\bibnamefont{Watrous}}, \bibinfo{journal}{J.
  Comp. System Sciences} \textbf{\bibinfo{volume}{62}}, \bibinfo{pages}{376}
  (\bibinfo{year}{2001}), \eprint{cs.CC/9812012}.

\bibitem[{\citenamefont{{Aharonov, D} et~al.}(2001)\citenamefont{{Aharonov, D},
  Ambainis, Kempe, and Vazirani}}]{aharonov00a}
\bibinfo{author}{\bibnamefont{{Aharonov, D}}},
  \bibinfo{author}{\bibfnamefont{A.}~\bibnamefont{Ambainis}},
  \bibinfo{author}{\bibfnamefont{J.}~\bibnamefont{Kempe}}, \bibnamefont{and}
  \bibinfo{author}{\bibfnamefont{U.}~\bibnamefont{Vazirani}}, in
  \emph{\bibinfo{booktitle}{Proc.~33rd Annual ACM STOC}}
  (\bibinfo{publisher}{ACM, NY}, \bibinfo{year}{2001}), pp.
  \bibinfo{pages}{50--59}, \eprint{quant-ph/0012090}.

\bibitem[{\citenamefont{Ambainis et~al.}(2001)\citenamefont{Ambainis, Bach,
  Nayak, Vishwanath, and Watrous}}]{ambainis01a}
\bibinfo{author}{\bibfnamefont{A.}~\bibnamefont{Ambainis}},
  \bibinfo{author}{\bibfnamefont{E.}~\bibnamefont{Bach}},
  \bibinfo{author}{\bibfnamefont{A.}~\bibnamefont{Nayak}},
  \bibinfo{author}{\bibfnamefont{A.}~\bibnamefont{Vishwanath}},
  \bibnamefont{and} \bibinfo{author}{\bibfnamefont{J.}~\bibnamefont{Watrous}},
  in \emph{\bibinfo{booktitle}{Proc.~33rd Annual ACM STOC}}
  (\bibinfo{publisher}{ACM, NY}, \bibinfo{year}{2001}), pp.
  \bibinfo{pages}{60--69}.

\bibitem[{\citenamefont{Farhi and Gutmann}(1998)}]{farhi98a}
\bibinfo{author}{\bibfnamefont{E.}~\bibnamefont{Farhi}} \bibnamefont{and}
  \bibinfo{author}{\bibfnamefont{S.}~\bibnamefont{Gutmann}},
  \bibinfo{journal}{Phys.~Rev.~A} \textbf{\bibinfo{volume}{58}},
  \bibinfo{pages}{915} (\bibinfo{year}{1998}), \eprint{quant-ph/9706062}.

\bibitem[{\citenamefont{Childs and Goldstone}(2004{\natexlab{a}})}]{childs03a}
\bibinfo{author}{\bibfnamefont{A.}~\bibnamefont{Childs}} \bibnamefont{and}
  \bibinfo{author}{\bibfnamefont{J.}~\bibnamefont{Goldstone}},
  \bibinfo{journal}{Phys.~Rev.~A} \textbf{\bibinfo{volume}{70}},
  \bibinfo{pages}{022314} (\bibinfo{year}{2004}{\natexlab{a}}),
  \eprint{quant-ph/0306054}.

\bibitem[{\citenamefont{Childs and Goldstone}(2004{\natexlab{b}})}]{childs04a}
\bibinfo{author}{\bibfnamefont{A.~M.} \bibnamefont{Childs}} \bibnamefont{and}
  \bibinfo{author}{\bibfnamefont{J.}~\bibnamefont{Goldstone}},
  \bibinfo{journal}{Phys.~Rev.~A} \textbf{\bibinfo{volume}{70}},
  \bibinfo{pages}{042312} (\bibinfo{year}{2004}{\natexlab{b}}),
  \eprint{quant-ph/0405120}.

\bibitem[{\citenamefont{Mackay et~al.}(2002)\citenamefont{Mackay, Bartlett,
  Stephenson, and Sanders}}]{mackay01a}
\bibinfo{author}{\bibfnamefont{T.~D.} \bibnamefont{Mackay}},
  \bibinfo{author}{\bibfnamefont{S.~D.} \bibnamefont{Bartlett}},
  \bibinfo{author}{\bibfnamefont{L.~T.} \bibnamefont{Stephenson}},
  \bibnamefont{and} \bibinfo{author}{\bibfnamefont{B.~C.}
  \bibnamefont{Sanders}}, \bibinfo{journal}{J.~Phys.~A: Math.~Gen.}
  \textbf{\bibinfo{volume}{35}}, \bibinfo{pages}{2745} (\bibinfo{year}{2002}),
  \eprint{quant-ph/0108004}.

\bibitem[{\citenamefont{Tregenna et~al.}(2003)\citenamefont{Tregenna, Flanagan,
  Maile, and Kendon}}]{tregenna03a}
\bibinfo{author}{\bibfnamefont{B.}~\bibnamefont{Tregenna}},
  \bibinfo{author}{\bibfnamefont{W.}~\bibnamefont{Flanagan}},
  \bibinfo{author}{\bibfnamefont{R.}~\bibnamefont{Maile}}, \bibnamefont{and}
  \bibinfo{author}{\bibfnamefont{V.}~\bibnamefont{Kendon}},
  \bibinfo{journal}{New J.~Phys.} \textbf{\bibinfo{volume}{5}},
  \bibinfo{pages}{83} (\bibinfo{year}{2003}), \eprint{quant-ph/0304204}.

\bibitem[{\citenamefont{Knight et~al.}(2003{\natexlab{a}})\citenamefont{Knight,
  Rold{\'a}n, and Sipe}}]{knight03a}
\bibinfo{author}{\bibfnamefont{P.~L.} \bibnamefont{Knight}},
  \bibinfo{author}{\bibfnamefont{E.}~\bibnamefont{Rold{\'a}n}},
  \bibnamefont{and} \bibinfo{author}{\bibfnamefont{J.~E.} \bibnamefont{Sipe}},
  \bibinfo{journal}{Phys.~Rev.~A} \textbf{\bibinfo{volume}{68}},
  \bibinfo{pages}{020301(R)} (\bibinfo{year}{2003}{\natexlab{a}}),
  \eprint{quant-ph/0304201}.

\bibitem[{\citenamefont{Knight et~al.}(2003{\natexlab{b}})\citenamefont{Knight,
  Rold{\'a}n, and Sipe}}]{knight03b}
\bibinfo{author}{\bibfnamefont{P.~L.} \bibnamefont{Knight}},
  \bibinfo{author}{\bibfnamefont{E.}~\bibnamefont{Rold{\'a}n}},
  \bibnamefont{and} \bibinfo{author}{\bibfnamefont{J.~E.} \bibnamefont{Sipe}},
  \bibinfo{journal}{Optics Comms.} \textbf{\bibinfo{volume}{227}},
  \bibinfo{pages}{147} (\bibinfo{year}{2003}{\natexlab{b}}),
  \eprint{quant-ph/0305165}.

\bibitem[{\citenamefont{Kendon and Sanders}(2004)}]{kendon04a}
\bibinfo{author}{\bibfnamefont{V.~M.} \bibnamefont{Kendon}} \bibnamefont{and}
  \bibinfo{author}{\bibfnamefont{B.~C.} \bibnamefont{Sanders}},
  \bibinfo{journal}{Phys.~Rev.~A} \textbf{\bibinfo{volume}{71}},
  \bibinfo{pages}{022307} (\bibinfo{year}{2004}), \eprint{quant-ph/0404043}.

\bibitem[{\citenamefont{Bach et~al.}(2004)\citenamefont{Bach, Coppersmith,
  Goldschen, Joynt, and Watrous}}]{bach02a}
\bibinfo{author}{\bibfnamefont{E.}~\bibnamefont{Bach}},
  \bibinfo{author}{\bibfnamefont{S.}~\bibnamefont{Coppersmith}},
  \bibinfo{author}{\bibfnamefont{M.~P.} \bibnamefont{Goldschen}},
  \bibinfo{author}{\bibfnamefont{R.}~\bibnamefont{Joynt}}, \bibnamefont{and}
  \bibinfo{author}{\bibfnamefont{J.}~\bibnamefont{Watrous}},
  \bibinfo{journal}{J.~Comput.~Syst.~Sci.} \textbf{\bibinfo{volume}{69}},
  \bibinfo{pages}{562} (\bibinfo{year}{2004}), \eprint{quant-ph/0207008}.

\bibitem[{\citenamefont{Annabestani}(2008)}]{annabestani08a}
\bibinfo{author}{\bibfnamefont{M.}~\bibnamefont{Annabestani}}
  (\bibinfo{year}{2008}), \bibinfo{note}{errors (now corrected) pointed out by
  private communication.}

\bibitem[{\citenamefont{Gattner}(2006)}]{gattner06a}
\bibinfo{author}{\bibfnamefont{A.}~\bibnamefont{Gattner}}
  (\bibinfo{year}{2006}), \bibinfo{note}{errors (now corrected) pointed out by
  private communication.}

\bibitem[{\citenamefont{Abal et~al.}(2006)\citenamefont{Abal, Siri, Romanelli,
  and Donangelo}}]{abal05a}
\bibinfo{author}{\bibfnamefont{G.}~\bibnamefont{Abal}},
  \bibinfo{author}{\bibfnamefont{R.}~\bibnamefont{Siri}},
  \bibinfo{author}{\bibfnamefont{A.}~\bibnamefont{Romanelli}},
  \bibnamefont{and}
  \bibinfo{author}{\bibfnamefont{R.}~\bibnamefont{Donangelo}},
  \bibinfo{journal}{Phys.~Rev.~A} \textbf{\bibinfo{volume}{73}},
  \bibinfo{pages}{069905(E)} (\bibinfo{year}{2006}), \bibinfo{note}{contains
  proof of limiting entanglement values.}, \eprint{ArXiv:quant-ph/0507264v4}.

\bibitem[{\citenamefont{Moore and Russell}(2002)}]{moore01a}
\bibinfo{author}{\bibfnamefont{C.}~\bibnamefont{Moore}} \bibnamefont{and}
  \bibinfo{author}{\bibfnamefont{A.}~\bibnamefont{Russell}}, in
  \emph{\bibinfo{booktitle}{Proc.~6th Intl.~Workshop on Randomization and
  Approximation Techniques in Computer Science (RANDOM '02)}}, edited by
  \bibinfo{editor}{\bibfnamefont{J.~D.~P.} \bibnamefont{Rolim}}
  \bibnamefont{and} \bibinfo{editor}{\bibfnamefont{S.}~\bibnamefont{Vadhan}}
  (\bibinfo{publisher}{Springer}, \bibinfo{year}{2002}), pp.
  \bibinfo{pages}{164--178}, \eprint{quant-ph/0104137}.

\bibitem[{\citenamefont{Kendon and Tregenna}(2003)}]{kendon02c}
\bibinfo{author}{\bibfnamefont{V.}~\bibnamefont{Kendon}} \bibnamefont{and}
  \bibinfo{author}{\bibfnamefont{B.}~\bibnamefont{Tregenna}},
  \bibinfo{journal}{Phys.~Rev.~A} \textbf{\bibinfo{volume}{67}},
  \bibinfo{pages}{042315} (\bibinfo{year}{2003}),
  \bibinfo{note}{arXiv:quant-ph/0209005}, \eprint{ArXiv:quant-ph/0209005}.

\bibitem[{\citenamefont{Ahmadi et~al.}(2003)\citenamefont{Ahmadi, Belk, Tamon,
  and Wendler}}]{ahmadi02a}
\bibinfo{author}{\bibfnamefont{A.}~\bibnamefont{Ahmadi}},
  \bibinfo{author}{\bibfnamefont{R.}~\bibnamefont{Belk}},
  \bibinfo{author}{\bibfnamefont{C.}~\bibnamefont{Tamon}}, \bibnamefont{and}
  \bibinfo{author}{\bibfnamefont{C.}~\bibnamefont{Wendler}},
  \bibinfo{journal}{Quantum Inform.~Compu.} \textbf{\bibinfo{volume}{3}},
  \bibinfo{pages}{611} (\bibinfo{year}{2003}), \eprint{quant-ph/0209106}.

\bibitem[{\citenamefont{Kempe}(2003{\natexlab{b}})}]{kempe02a}
\bibinfo{author}{\bibfnamefont{J.}~\bibnamefont{Kempe}}, in
  \emph{\bibinfo{booktitle}{Proc.~7th Intl.~Workshop on Randomization and
  Approximation Techniques in Computer Science (RANDOM '03)}}
  (\bibinfo{publisher}{Springer, Heidelberg},
  \bibinfo{year}{2003}{\natexlab{b}}), Lecture Notes in Computer Science, pp.
  \bibinfo{pages}{354--369}, \eprint{quant-ph/0205083}.

\bibitem[{\citenamefont{Watrous}(2002)}]{watrous02a}
\bibinfo{author}{\bibfnamefont{J.}~\bibnamefont{Watrous}}
  (\bibinfo{year}{2002}), \bibinfo{note}{private communication}.

\end{thebibliography}
%--------------------%

%\vspace{2em}

%\fbox{
%\large
%This paper is online as \texttt{quant-ph/0504042}
%}

%----------------------%
%%% must end with... %%%
\end{document}